# A CONTROL CO-DESIGN FRAMEWORK TO OPTIMIZE SUSTAINABILITY WITH APPLICATION TO MICROGRID-DRIVEN DATA CENTERS

**Tania Rifat Jahan[1], Donald J. Docimo[1,*]**

[1]Department of Mechanical & Aerospace Engineering, Texas Tech University, Lubbock, TX 79409, USA

*Corresponding Author's E-mail: donald.docimo@ttu.edu

## ABSTRACT

*This work studies the optimization of physical plant characteristics and controller parameters for sustainability. Environmental sustainability is strongly correlated with the development and operation of energy systems, with data centers as the preeminent modern example. Data centers, and the grid technologies that provide their power, consume nonnegligible amounts of global energy and pose a risk to increase greenhouse gas (GHG) emissions and electronic waste. Addressing such issues requires improved plant design and control strategies, often approached through optimization-based methods. However, there are noticeable gaps regarding data center and microgrid design: (i) simultaneous optimization of plant and controller features is rarely explored, and (ii) sustainability criteria are not emphasized. This work addresses these gaps by establishing a generalized, sustainability-centric control co-design (CCD) framework for energy systems. Uniquely, the CCD framework defines and categorizes sustainability metrics into three lifecycle stages - manufacturing, operation, and disposal - supporting optimization and comparative analysis of the metrics for different design options. To exemplify its use, the CCD framework is applied to a microgrid-driven data center system, providing a family of sustainability metrics correlated to plant and controller parameters. The analysis enabled by the framework supports the determination of tradeoffs between different types of environmental sustainability metrics. The system designs identified by the proposed framework show substantial improvements to environmental sustainability categories as compared to designs identified through baseline procedures.*

## 1. INTRODUCTION

Environmental and economic sustainability is invariably linked to two aspects of energy systems: their *physical design* and their *control*. This is exemplified most by data centers, being the key load for modern grids and microgrids. Data centers have been recorded to consume 2% of the world's net generated electricity per year [1], with this percentage currently skyrocketing due to the rapid emergence of artificial intelligence (AI) and cloud users [2]. Increasingly, microgrids are developed for the singular purpose of powering these data centers [3]. With the prevalence of microgrids and data centers comes mandates to improve energy efficiency and environmental sustainability. These improvements are impeded by the systems' complexity, requiring coordination of numerous components with multi-domain and multi-timescale dynamics. Prior research on the advancement of microgrids, data centers, and stationary energy systems can be categorized into two groups: the broadening of the design space, and determining pertinent objective functions.

## 1.1 Broadening the Design Space

System-level performance can be strengthened by searching through a larger design space, often accomplished by incorporating a broader set of design options. This area of the literature includes the use of design variables or alternative designs to represent nontraditional plant features. Sizing design variables can describe photovoltaic (PV) and battery components in microgrid optimization problems [4–7], solvable through methods such as mixed-integer linear programming [8,9]. Similarly, plenum depth, ceiling height, and tile placement in data centers have been explored in an optimization context [10]. Alternative designs can describe unconventional aisle containment configurations for data centers [11], analyzable through ANSYS CFX software [12] or even experiments [13]. This area of the literature also includes the use of design variables to capture controller parameters [14–16], using techniques such as particle swarm optimization [17]. A limited number of studies pursue the simultaneous optimization of both plant and controller parameters, a process known as control co-design (CCD). CCD approaches have been applied for microgrids to increase robustness with respect to attacks [18], reliability [19], and integration of hybrid tidal-solar-battery components [20].

## 1.2 Determining Pertinent Objectives

System-level performance can also be strengthened by defining, identifying, or otherwise utilizing objective functions that properly reflect system goals. Traditional objectives for energy systems cover broad classifications such as regulation, cost, and functionality. Regulation objectives are similar to state tracking and constraint obedience, including voltage and frequency regulation [21] for microgrids and temperature and coolant flow rate control in data centers [22,23]. Cost can be represented through different objective functions. Operational costs, as well as related energy costs, are often defined for controllers to minimize. This can involve determination of how to operate PVs, energy storage, and loads [24–26], as well as cooling elements [27]. Purchase costs can be related to plant component sizes [28]. Functionality-based objectives are wide-ranging and relate to the overall system behavior. Examples include optimizing and monitoring cooling efficiency in data centers [29], and mitigating battery degradation [30]. Less conventional studies balance economic benefits against minimization of greenhouse gas (GHG) emissions [30,31].

## 1.3 Gaps and Contributions

Despite the progress made with respect to design and control of microgrids, data centers, and related energy systems, gaps in the literature persist. First, there are limited studies on CCD of data centers and microgrids that drive data centers. Second, few studies quantify sustainability [32,33], regulating sustainability criteria to a small part of optimization problems, rather than the focus. If unaddressed, environmental impacts from these systems will remain unmitigated. Built atop the authors' previous study on microgrid sustainability [34], this article fills the gaps through the following contributions:

1. A sustainability-centric CCD framework for energy system optimization is developed. The framework introduces sustainability metrics as the primary objective functions, facilitated by categorization of the metrics across three component lifecycle stages: manufacturing, operation, and disposal.

2. A novel, CCD-oriented representation of a microgrid-powered data center is developed. The proposed framework is applied to this model, yielding a family of correlations from plant and controller parameters to sustainability metrics.
3. Via the framework, insights into the relative relationships among sustainability objective functions, as well as Pareto-optimal designs, are determined.

To cover these contributions, Section 2 begins by developing the CCD optimization framework for energy systems to incorporate sustainability metrics for component lifecycle stages, which are thoroughly defined. Section 3 outlines the CCD-oriented candidate model of the microgrid and data center. Section 4 presents the controller used to manage the system, validating performance for a small set of designs. Section 5 applies the sustainability-centric CCD framework to the candidate system for a 24-hour blackout where only the microgrid can provide power to the data center. Section 6 analyzes different relationships between the design variables and objective functions. It also presents the Pareto-optimal solutions and compares the proposed framework with a baseline case. Section 7 provides a synopsis of the article's findings and concludes with some future directions.

## 2. SUSTAINABILITY-CENTRIC CONTROL CO-DESIGN FRAMEWORK

This section presents the proposed CCD framework for sustainability-centric optimization of energy systems. The framework is divided into five steps. The first three steps define key elements for optimization: the critical design variables, the objective functions, and the constraints. The fourth step constructs the multi-objective optimization problem using these elements. The fifth step solves the problem, identifies optimal designs, and analyzes the outcomes.

### 2.1 Define the Design Variables

The design variables characterize the system's adjustability, with the variable values determined through an optimization algorithm. To expand the design space, the proposed framework incorporates both plant and controller design variables. The plant design variables, denoted as $\boldsymbol{\theta} = \{\theta_1, \theta_2, \dots, \theta_{N_\theta}\}$, can include both sizing- and topology-related parameters of the system. Examples include PV array configuration, number of battery cells, and converter size and material selection. The controller design variables, denoted as $\boldsymbol{\phi} = \{\phi_1, \phi_2, \dots, \phi_{N_\phi}\}$, can include control architecture settings, timesteps, predictive horizons, and gains within a control algorithm. The values $N_\theta$ and $N_\phi$ indicate the total number of plant and controller design variables, respectively.

### 2.2 Define the Objective Functions

A novel aspect of this CCD framework is its ability to highlight and integrate quantified measures of sustainability for optimization. Shaping this as a multi-objective problem, environmental sustainability objectives are grouped into three categories of metrics:

- *Manufacturing:* The objective function $J_m$ captures all sustainability metrics relating to the manufacturing of components for, and the construction of, the energy system. This encompasses potential environmental impacts resulting from net energy consumption, waste generation, and pollutants emitted during the extraction, transportation, and assembly of raw materials. The purpose of this objective is to describe the early lifecycle stage of the system. An example includes quantifying the emissions produced from manufacturing solar panels, which often involves energy consumption from traditional, GHG-emitting energy production [35].

- *Operation:* The objective function $J_o$ accounts for environmental impacts relating to the operation of the energy system. This includes impacts from electricity consumption, resource use, control strategies, heat generation, and noise production. This objective's purpose is to capture sustainability metrics during the middle lifecycle stage of the system. One example is the impact of waste heat generated on the nearby environment.
- *Disposal:* The objective function $J_d$ agglomerates sustainability metrics categorized by the ending lifecycle stage for plant components, electronic devices, and other pieces of the system. Devices that are not recycled are disposed of at the end of their life as electronic waste (e-waste). The disposal objective function refers to the environmental impacts resulting from improper waste disposal, inadequate recycling, and increased demand for new resources.

The terms $J_m, J_o$, and $J_d$ are all functions of the design variables, $\boldsymbol{\theta}$ and $\boldsymbol{\phi}$. The individual objective functions can be incorporated into a total objective function, $J_{tot}$, with $w_m$, $w_o$, and $w_d$ as weights associated with each metric:

$$J_{tot} = w_m {J_m}^N + w_o {J_o}^N + w_d {J_d}^N \tag{1}$$

In Equation (1), the parameter $N \geq 1$ can be set to larger values to implement compromise programming [36], which can be desirable for nonconvex problems. The categorization of sustainability metrics supports comparative analysis, permitting identification of the primary lifecycle stage that creates negative sustainability impacts. This has the potential to inform selection of designs along the Pareto front.

### 2.3 Define the Constraints

The constraints are used to ensure the design process obeys physical and practical limitations. The most prominent set of equality constraints captures the dynamics, including how components physically interact with one another. For the energy systems considered, the dynamics are assumed to be coupled ordinary differential and algebraic equations (DAEs). Equation (2) presents the representation of the plant dynamics at a time $t$:

$$\boldsymbol{E}(\boldsymbol{\theta})\,\dot{\boldsymbol{x}}(t) - \boldsymbol{f_P}(\boldsymbol{x}(t), \boldsymbol{u}(t), \boldsymbol{d}(t), \boldsymbol{\theta}) = \boldsymbol{0} \tag{2}$$

The state vector $\boldsymbol{x}$, of size $N_x \times 1$, serves to describe changes in the underlying dynamics. The control input vector $\boldsymbol{u}$, of size $N_u \times 1$, describes the variables manipulated by the online management algorithm. The exogeneous input vector $\boldsymbol{d}$, of size $N_d \times 1$, are known time-varying trajectories that include effects at the system boundary, and reference signals such as desired state values to track. Along with $\boldsymbol{\theta}$, these feed into the nonlinear plant functions within the $N_x$-sized $\boldsymbol{f_P}$. The square mass matrix $\boldsymbol{E}$, which is a function of the plant design variables, is potentially singular. Equation (3) presents the constraint regarding the initial conditions for the states, given as $\boldsymbol{x_{IC}}$. The initial conditions are defined such that all algebraic terms in Equation (2) are satisfied at $t = 0$.

$$\boldsymbol{x}(0) - \boldsymbol{x_{IC}} = \boldsymbol{0} \tag{3}$$

Related to the plant dynamics are equality constraints to determine the control input values, as presented in Equation (4). The control inputs are determined from functions within the $N_u$-sized $\boldsymbol{f_C}$ that depend on the plant and controller design variables, as well as the most recent state and exogeneous input values. The dependency can extend to prior values of $\boldsymbol{x}$ and $\boldsymbol{u}$, denoted by the $t$ dependency shown within $\boldsymbol{f_C}$ in Equation (4). This formulation keeps the CCD framework modular with respect to a variety of control algorithms. Examples include PI algorithms [37], model-free methods [34], and nested optimization problems for the controller [38,39].

$$\boldsymbol{u}(t) - \boldsymbol{f_C}(\boldsymbol{x}(t), \boldsymbol{d}(t), \boldsymbol{\theta}, \boldsymbol{\phi}) = \boldsymbol{0} \tag{4}$$

The inequality constraints for the optimization problem are used to accomplish two things. The first set of inequality constraints are to ensure traditional metrics of success are met by the designed system. These metrics, defined by the functions that make up $\boldsymbol{\mu}$, are objective functions to be minimized in baseline approaches. In this work, these are instead constrained by upper bound $\boldsymbol{\mu_{max}}$, the largest acceptable value:

$$\boldsymbol{\mu}(t, \boldsymbol{x}, \boldsymbol{u}, \boldsymbol{d}, \boldsymbol{\theta}, \boldsymbol{\phi}) \leq \boldsymbol{\mu_{max}} \tag{5}$$

As examples, $\boldsymbol{\mu}$ can contain functions describing state tracking error or losses in energy efficiency over a given test trajectory. The $\boldsymbol{\mu_{max}}$ values would then be the maximum acceptable error or reduction in efficiency.

The second set of inequality constraints for the optimization problem reflects thresholds in the plant and controller design variable values. Equation (6) defines these, with $\boldsymbol{\theta_{min}}$ and $\boldsymbol{\theta_{max}}$ as the minimum and maximum values of $\boldsymbol{\theta}$, respectively. Similarly, $\boldsymbol{\phi_{min}}$ and $\boldsymbol{\phi_{max}}$ are the minimum and maximum values of $\boldsymbol{\phi}$, respectively.

$$\begin{gathered} \boldsymbol{\theta_{min}} \leq \boldsymbol{\theta} \leq \boldsymbol{\theta_{max}} \\ \boldsymbol{\phi_{min}} \leq \boldsymbol{\phi} \leq \boldsymbol{\phi_{max}} \end{gathered} \tag{6}$$

### 2.4 Formulate the Optimization Problem

Based on the previous three subsections, this step formulates the CCD problem. Equation (7) presents the full optimization problem. To ease presentation, this is structured to consider one predefined $\boldsymbol{d}$ profile over a time period from zero to final time $t_f$, with known initial conditions $\boldsymbol{x_{IC}}$. However, alternative or multiple profiles can be incorporated with only minor changes to the formulation.

$$\begin{aligned} \min_{\boldsymbol{\theta}, \boldsymbol{\phi}} \quad & J_{tot} = w_m {J_m}^N + w_o {J_o}^N + w_d {J_d}^N \\ \text{s.t.:} \quad & \boldsymbol{E}(\boldsymbol{\theta})\dot{\boldsymbol{x}}(t) - \boldsymbol{f_P}(\boldsymbol{x}(t), \boldsymbol{u}(t), \boldsymbol{d}(t), \boldsymbol{\theta}) = \boldsymbol{0}, \\ & \qquad \forall\, t \in [0, t_f] \\ & \boldsymbol{u}(t) - \boldsymbol{f_C}(\boldsymbol{x}(t), \boldsymbol{d}(t), \boldsymbol{\theta}, \boldsymbol{\phi}) = \boldsymbol{0},\ \forall\, t \in [0, t_f] \\ & \boldsymbol{x}(0) - \boldsymbol{x_{IC}} = \boldsymbol{0} \\ & \boldsymbol{\mu}(t, \boldsymbol{x}, \boldsymbol{u}, \boldsymbol{d}, \boldsymbol{\theta}, \boldsymbol{\phi}) \leq \bar{\boldsymbol{\mu}} \\ & \boldsymbol{\theta}_{min} \leq \boldsymbol{\theta} \leq \boldsymbol{\theta}_{max} \\ & \boldsymbol{\phi}_{min} \leq \boldsymbol{\phi} \leq \boldsymbol{\phi}_{max} \end{aligned} \tag{7}$$

There are several options for completing a single function evaluation for a set of $\boldsymbol{\theta}$ and $\boldsymbol{\phi}$ values. The shooting method evaluates the equality constraints to determine $\boldsymbol{x}$ and $\boldsymbol{u}$ trajectories over the time period [40]. This can be accomplished through numerical discretization or exact methods if available. Alternatively, direct transcription adds $\boldsymbol{x}$ and $\boldsymbol{u}$ to the list of decision variables to determine those trajectories simultaneously with optimal $\boldsymbol{\theta}$ and $\boldsymbol{\phi}$ values.

### 2.5 Solve and Analyze

The optimal values of $\boldsymbol{\theta}$ and $\boldsymbol{\phi}$ are determinable through a number of solution methods: grid search, gradient descent, and genetic algorithms are a representative few. However, the purpose of the framework is not only to solve Equation (7), but also to support analysis of sustainability in a CCD context. The following sensitivity analyses are proposed to better understand sustainability-centric CCD, based on the division of sustainability metrics into the three categories:

- Determine the sensitivity of each sustainability category's objective against the design variables: $\frac{\partial J_m}{\partial \boldsymbol{\theta}}$, $\frac{\partial J_o}{\partial \boldsymbol{\theta}}$, $\frac{\partial J_d}{\partial \boldsymbol{\theta}}$, $\frac{\partial J_m}{\partial \boldsymbol{\phi}}$, $\frac{\partial J_o}{\partial \boldsymbol{\phi}}$, and $\frac{\partial J_d}{\partial \boldsymbol{\phi}}$. Comparison of these terms will (1) inform which design variable should be the focus to improve sustainability. The comparison will also (2) help determine whether plant or controller design variables have a greater influence on the objectives.
- Compare the sensitivities of the different categories of objectives with respect to each other (e.g., $\frac{\partial J_m}{\partial J_o}$, $\frac{\partial J_m}{\partial J_d}$, $\frac{\partial J_o}{\partial J_d}$). By looking at these terms along the Pareto front, the designer will be able to identify conflicts and make decisions based on their priorities. This knowledge will facilitate prioritization of one or more sustainability categories.

## 3. MODEL DYNAMICS

This section outlines the microgrid and data center component models that are combined to build the full system representation. Shown in Figure 1, the microgrid serves as the energy-producing half of the system, with renewable generation and battery storage components. The data center acts as the energy-consuming half, with demands to power servers and cooling subsystems. The parameters of the system are drawn from numerical and experimental studies in the literature [12,41–46] and are detailed in Appendix A.

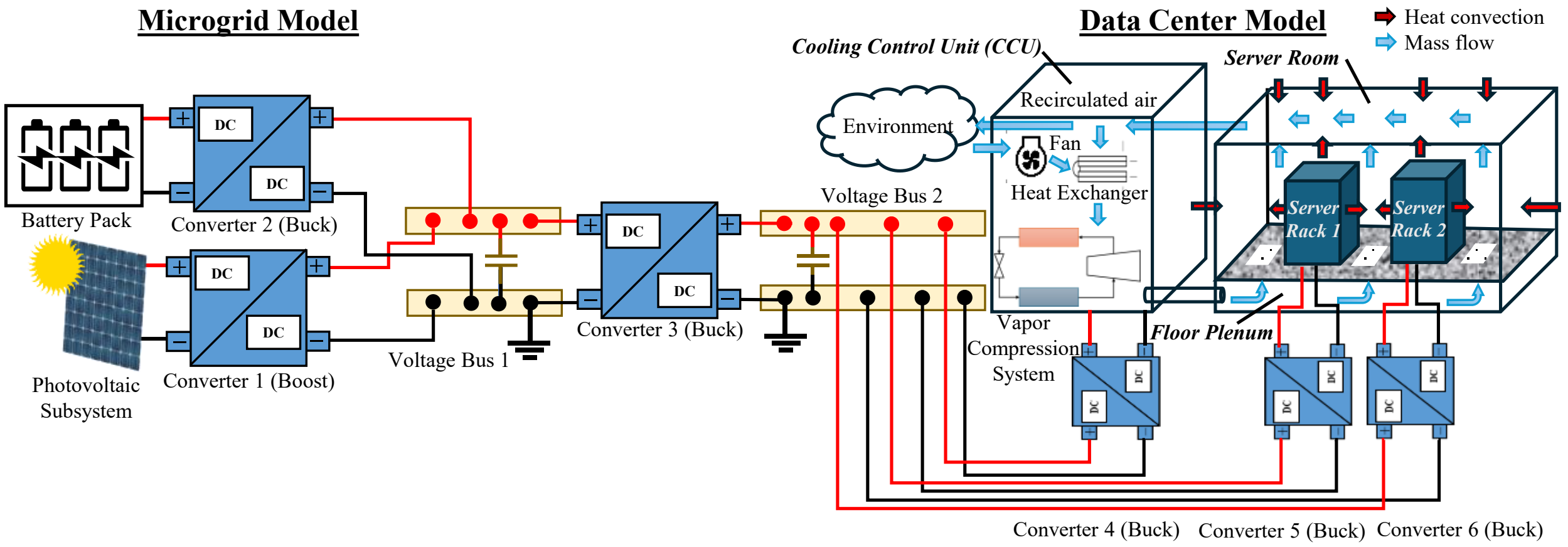

**FIGURE 1:** CONFIGURATION OF MIGROGRID MODEL WITH DATA CENTER AS THE LOAD

## 3.1 Microgrid Components

There are five types of components used to model microgrid dynamics: (i) voltage bus, (ii) boost converter, (iii) buck converter, (iv) PV subsystem, (v) battery pack. The electrical component models are based on those found in the literature [47,48], but tuned for the data center-related applications explored in this work.

### *3.1.1 Voltage Bus*

The two voltage buses act as hubs, providing connection points for all power converters. Voltage bus 1 supports connection to the energy generation and storage components, and voltage bus 2 serves as the bridge between the microgrid and the data center. Up to four connected current sources or loads are captured by $I_{bus,1}$ to $I_{bus,4}$ for each bus, shown in Figure 2. With electrical capacitance $C_{bus}$ and resistances $R_{b,1}$ and $R_{b,2}$ to capture inefficiencies, voltage state $V_{bus}$ for $i = [1,2]$ has dynamics described by Equations (8).

$$\dot{V}_{bus,i} = \frac{1}{C_{bus}}\left(I_{bus,1,i} + I_{bus,2,i} + I_{bus,3,i} + I_{bus,4,i} - \frac{V_{bus,i}}{R_{b,1}}\right) \tag{8}$$

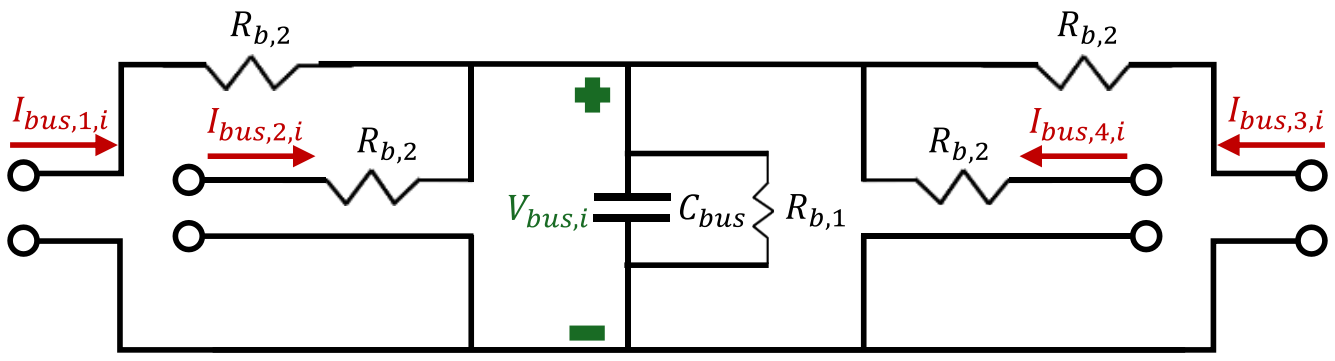

**FIGURE 2:** CIRCUIT DIAGRAM FOR VOLTAGE BUS $i$

### *3.1.2 Boost Converter*

Converter 1 is a boost converter that connects the PV subsystem to voltage bus 1. It has two electrical capacitances of value $C$ that relate to an input voltage ($V_{in}$) and output voltage ($V_{out}$), shown in Figure 3(a). It also has inefficiencies modeled by resistances $R_{c,1}$ and $R_{c,2,j}$, as well as a controllable duty cycle $D$. The dynamics are described by Equation (9), with $I_{in}$ and $I_{out}$ as the input and output currents of the converter, respectively. For the booster converter, $j = 1$.

$$\begin{aligned} \dot{V}_{in,j} &= \frac{1}{C}\left(I_{in,j} - \frac{V_{in,j}}{R_{c,1}} - \frac{1}{R_{c,2,j}}\left(V_{in,j} - D_j V_{out,j}\right)\right) \\ \dot{V}_{out,j} &= \frac{1}{C}\left(\frac{D_j}{R_{c,2,j}}\left(V_{in,j} - D_j V_{out,j}\right) - \frac{V_{out,j}}{R_{c,1}} - I_{out,j}\right) \end{aligned} \tag{9}$$

### *3.1.3 Buck Converter*

The remaining converters are buck DC-DC converters, shown in Figure 3(b). Converter 2 connects the Li-ion battery pack to voltage bus 1, and converter 3 connects the two buses. The other converters are used to control the electrical power supplied to the cooling system, server rack 1, and server rack 2. Equation (10) summarizes the dynamics, with parameters defined in the same manner as the boost converter, for $j = [2..6]$.

$$\dot{V}_{in,j} = \frac{1}{C}\left(I_{in,j} - \frac{V_{in,j}}{R_{c,1}} - \frac{D_j}{R_{c,2,j}}\left(D_j V_{in,j} - V_{out,j}\right)\right)$$
$$\dot{V}_{out,j} = \frac{1}{C}\left(\frac{1}{R_{c,2,j}}\left(D_j V_{in,j} - V_{out,j}\right) - \frac{V_{out,j}}{R_{c,1}} - I_{out,j}\right) \tag{10}$$

### *3.1.4 PV Subsystem*

In this system, the PV provides renewable energy and supplies the required power to the data center. Equation (11) describes how the temperature $T_{PV}$ is affected by irradiance $G$, ambient temperature $T_{\infty}$, and electrical power output $I_{PV}V_{PV}$. The other parameters are the lumped thermal capacitance $C_{PV}$, absorptivity $\alpha_{PV}$, module surface area $A_{PV}$, and convection coefficient $h_{PV}$. The current $I_{PV}$ is represented as a current source with nonlinear dependencies on $T_{PV}$, $G$, and $V_{PV}$ [49]. A total of $N_{par,PV}$ solar panels are placed in parallel.

$$\dot{T}_{PV} = \frac{1}{C_{PV}}\left(\alpha_{PV} A_{PV} G - h_{PV} A_{PV}(T_{PV} - T_{\infty}) - I_{PV}V_{PV}\right) \tag{11}$$

### *3.1.5 Battery Pack*

The Li-ion battery pack acts as the energy storage device, accumulating excess energy and releasing it when needed to meet the load. The equivalent circuit model of Figure 4 describes a battery cell, with dynamics described through Equations (12)-(14). The states of the battery are the state of charge ($SOC$) and relaxation voltages $V_{re,k}$ for $k \in [2,3,4]$, with total battery voltage $V_{Li}$. Battery current $I_{Li}$ is determinable using Equation (14). Table A.1 in Appendix A presents values for resistances $R_{Li,1}$ to $R_{Li,4}$, capacitances $C_{Li,2}$ to $C_{Li,4}$, and capacity $Q_{Li}$. Also provided is the nonlinear dependency of the open-circuit voltage $OCV_{Li}$ on $SOC$. $N_{par,Li}$ cells are grouped in parallel to make modules, and $N_{ser,Li}$ modules are connected in series to form the pack. The number of batteries connected in series is specified to meet a standardized voltage for bus 1. The number of cells in parallel is used to modulate the pack's energy capacity, and will serve as a design variable in when applying the proposed CCD framework.

$$\dot{SOC} = -\frac{I_{Li}}{N_{par,Li} Q_{Li}} \tag{12}$$

$$\dot{V}_{re,k} = \frac{1}{C_{Li,k}}\left(\frac{I_{Li}}{N_{par,Li}} - \frac{V_{re,k}}{R_{Li,k}}\right) \tag{13}$$

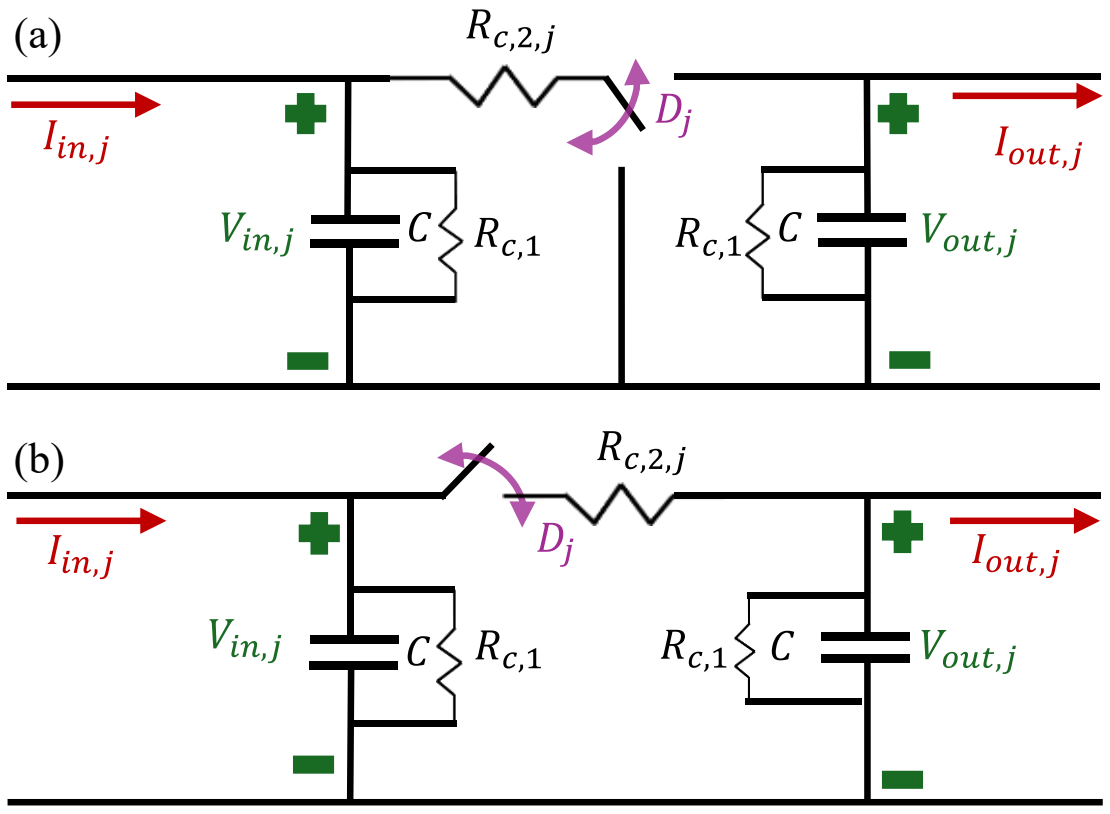

**FIGURE 3:** CIRCUIT DIAGRAM FOR A (a) BOOST OR (b) BUCK CONVERTER *j*

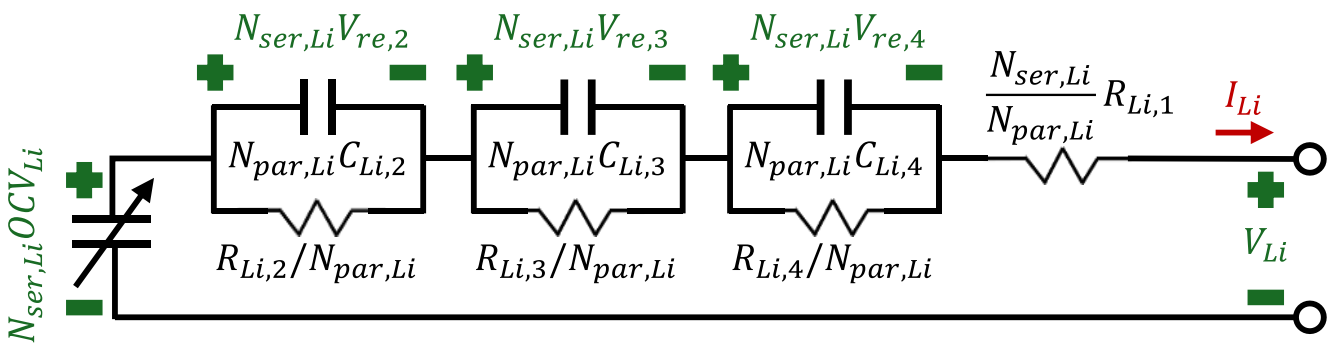

**FIGURE 4:** CIRCUIT DIAGRAM FOR A LI-ION BATTERY PACK

$$V_{Li} = N_{ser,Li}\left(OCV_{Li}(SOC) - V_{re,2} - V_{re,3} - V_{re,4} - I_{Li}\frac{R_{Li,1}}{N_{par,Li}}\right) \quad (14)$$

### 3.2 Data Center Components

A data center is a custom-designed facility that houses computing infrastructure, such as servers, for data storage and computation. Figure 1 presents the major sectors of the data center. Air is circulated through the building for cooling. The cooling control unit (CCU) draws in hot air from the server room, expelling a fraction ($\alpha$) and recirculating the rest. The fan drives the recirculated air, combined with ambient air drawn in, into the heat exchanger of the vapor compression system (VCS). This cools the air before it enters the plenum and server room, with fractions $\beta$ and $\gamma$ capturing the split of air. There are five types of dynamic models used to describe components and elements of the data center.

#### *3.2.1 Server Rack*

The servers are stacked on top of one another within each rack, separated by air gaps, and modeled as a single control volume (CV) with a lumped thermal capacitance $C_s$ shown in Figure 5. Equation (15) represents the temperature $T_s$ dynamics of server racks $i = [1,2]$.

$$\dot{T}_{s,i} = \frac{1}{C_s}\left(P_{s,i} - hA_{s,1}\left(T_{s,i} - \mathbb{T}_{a,l,i}\right) - hA_{s,1}\left(T_{s,i} - \mathbb{T}_{a,r,i}\right) - hA_{s,2}\left(T_{s,i} - \mathbb{T}_{a,c,i}\right)\right) \quad (15)$$

The first term is input electrical power coming from a converter to the rack, $P_s$, which converts to heat during operation. The remaining three terms describe convective heat transfer. An effective convection coefficient $h$ is used to capture transfer effects from air flowing next to and through the server rack as a lumped term. Convection on the sides of the rack acts through surface area $A_{s,1} = h_s t_s$, with $h_s$ as the rack height and $t_s$ as its depth or thickness. The rack height is proportional to the number of servers per rack. This will be used as a design variable in the application of proposed

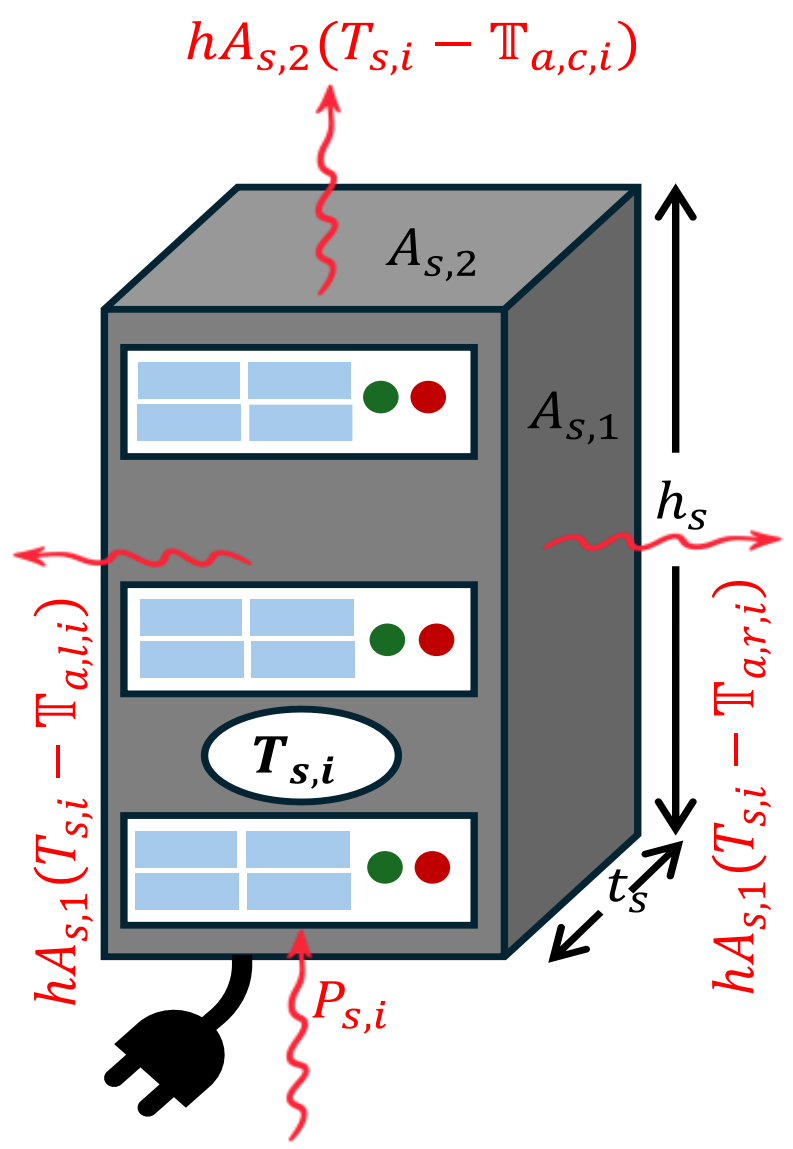

**FIGURE 5:** SCHEMATIC DIAGRAM OF A SERVER RACK $i$

CCD framework. The air temperatures to the left and right of the rack are $\mathbb{T}_{a,l}$ and $\mathbb{T}_{a,r}$, respectively. Convection also acts through the ceiling of the rack, with surface area $A_{s,2}$ and air temperature $\mathbb{T}_{a,c}$ at the ceiling.

### *3.2.2 Server Room Air*

The $j = [1..8]$ CVs of air that surround the racks are used to represent air circulation in the server room, shown in Figure 6. Equation (16) presents the generalized form of a CV's dynamics, with thermal capacitance $C_{sr}$ and air specific heat $c_p$ approximated as constant. There are up to five power flows that can influence the CV's temperature $T_{sr}$. The first two are tied to incoming air at mass flow rates $\dot{m}_{sr,1}$ and $\dot{m}_{sr,2}$, with temperatures $\mathbb{T}_{sr,1}$ and $\mathbb{T}_{sr,2}$, respectively. The third is the rate of energy exiting from air moving at mass flow rate $\dot{m}_{sr,3}$. The fourth and fifth are both used to capture convective heat transfer through convection coefficient $h$, surface areas $A_{sr,1}$ and $A_{sr,2}$, and external temperatures $\mathbb{T}_{sr,3}$ and $\mathbb{T}_{sr,4}$.

$$\dot{T}_{sr,j} = \frac{1}{C_{sr,j}}\Big(\dot{m}_{sr,1,j}c_p\mathbb{T}_{sr,1,j} + \dot{m}_{sr,2,j}c_p\mathbb{T}_{sr,2,j} - \dot{m}_{sr,3,j}c_pT_{sr,j} + hA_{sr,1,j}\big(\mathbb{T}_{sr,3,j} - T_{sr,j}\big) + hA_{sr,2,j}\big(\mathbb{T}_{sr,4,j} - T_{sr,j}\big)\Big) \tag{16}$$

### *3.2.3 Cooling Control Unit Air*

Similar to server room air, CCU air is divided into $k = [1..4]$ CVs. The CCU is composed of interconnected thermal CVs that regulate air recirculation, mixing, cooling, and rejection of heat to the environment. This includes: (i) a volume that partially rejects air from the building, recirculating the rest; (ii) cool environment air driven by a fan, providing energy to propel air forward; (iii) mixed recirculated and environment air; (iv) air cooled by the VCS. Equation (17) presents the general state equation for air temperature $T_{ccu}$, with thermal capacitance $C_{ccu}$. There are five potential power flows for a CCU CV. Two power flows are from air entering at mass flow rates $\dot{m}_{ccu,1}$ and $\dot{m}_{ccu,2}$ with temperatures $\mathbb{T}_{ccu,1}$ and $\mathbb{T}_{ccu,2}$. Two power flows are from air exiting at mass flow rates $\dot{m}_{ccu,3}$ and $\dot{m}_{ccu,4}$ with temperature $T_{ccu}$. One power flow, $P_{cool}$, captures a cooling effect from an external device like a VCS. Figure 7 shows a representative CV for cooling control unit air.

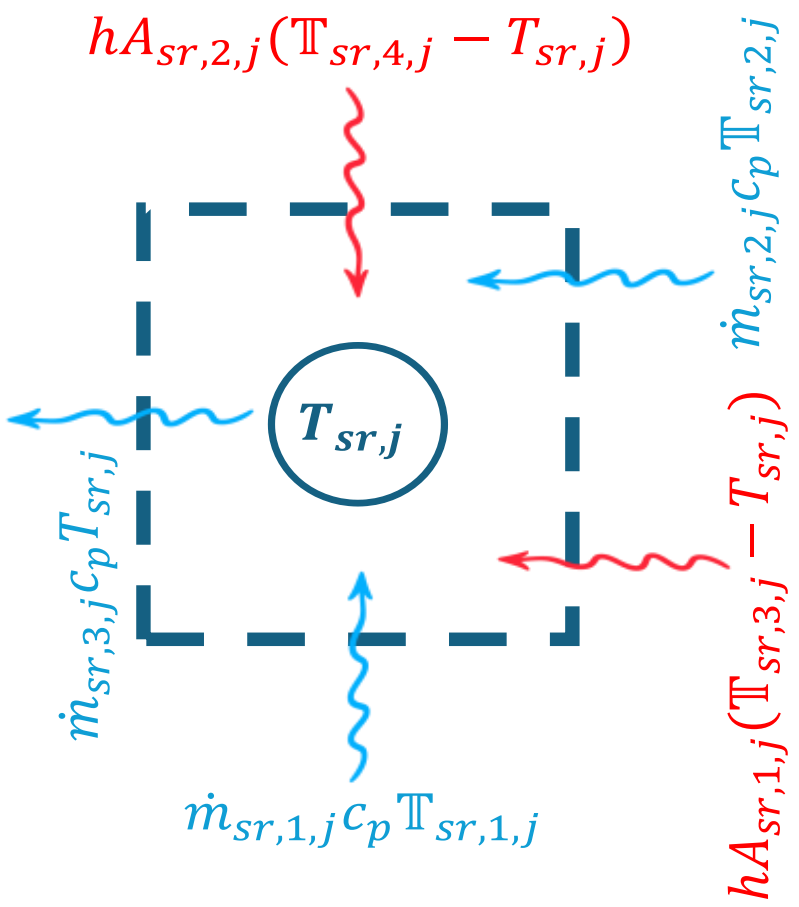

**FIGURE 6:** SCHEMATIC DIAGRAM FOR A CV SERVER ROOM AIR *j*

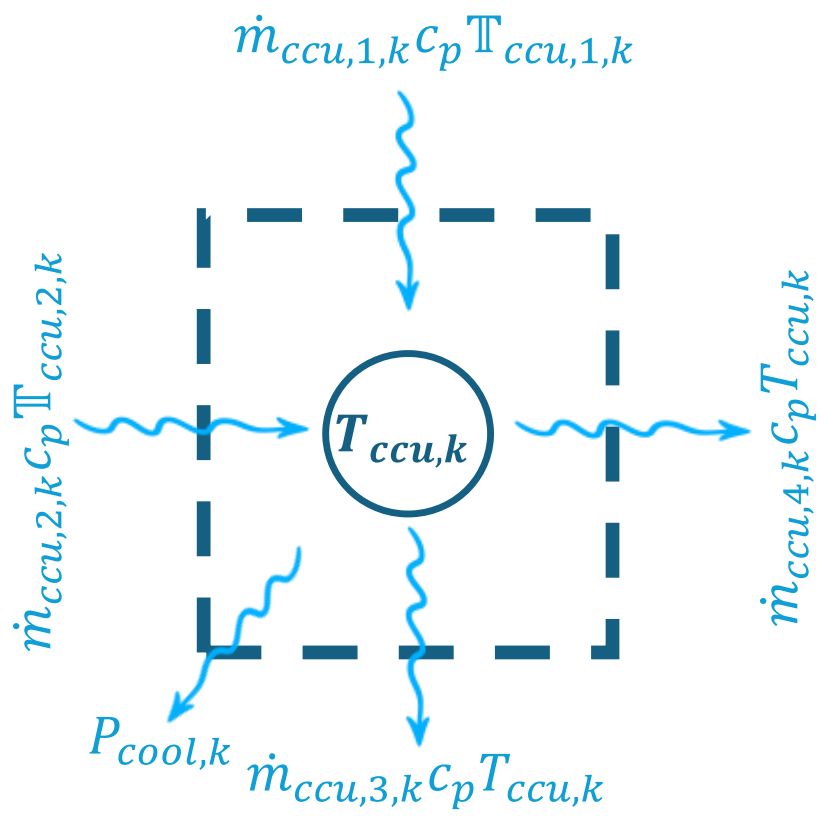

**FIGURE 7:** SCHEMATIC DIAGRAM FOR A CV COOLING CONTROL UNIT AIR *k*

$$\dot{T}_{ccu,k} = \frac{1}{C_{ccu,k}}\left(\dot{m}_{ccu,1,k}c_p\mathbb{T}_{ccu,1,k} + \dot{m}_{ccu,2,k}c_p\mathbb{T}_{ccu,2,k} - \dot{m}_{ccu,3,k}c_pT_{ccu,k} - \dot{m}_{ccu,4,k}c_pT_{ccu,k} - P_{cool,k}\right) \quad (17)$$

### *3.2.4 Plenum Air*

A plenum is a raised floor under the server room, allowing cool air from the CCU to be passed through the floor tiles to the racks [22]. Equation (18) presents the dynamic for the $n = [1..3]$ plenum air temperatures, $T_p$, with thermal capacitances $C_p$. The variables $\dot{m}_{p,1}$, $\dot{m}_{p,2}$, and $\dot{m}_{p,3}$ are mass flow rates of air in and out of the CV. Temperature $\mathbb{T}_{p,1}$ is that which enters the plenum section.

$$\dot{T}_{p,n} = \frac{1}{C_{p,n}}\left(\dot{m}_{p,1,n}c_p\mathbb{T}_{p,1,n} - \dot{m}_{p,2,n}c_pT_{p,n} - \dot{m}_{p,3,n}c_pT_{p,n}\right) \quad (18)$$

### *3.2.5 Vapor Compression System*

The VCS interacts with air through a heat exchanger, cooling it down as the fluid passes through. Equation (19) presents a first-order dynamic for the VCS [48], with state $X_{vcs}$ and capacitance $C_{vcs}$. The variable $P_{vcs}$ is the electrical power put into the VCS, with the coefficient of performance ($COP$) as a nonlinear function of the air temperature. The term $X_{vcs}COP$ indicates the amount of heat removed from the air.

$$\dot{X}_{vcs} = \frac{1}{C_{vcs}}(P_{vcs} + X_{vcs}COP - (1 + COP)X_{vcs}) \quad (19)$$

## 3.3 Combined Microgrid and Data Center Model

Development of the CCD-oriented, microgrid-driven data center model is accomplished by coupling the individual component models. The first stage of this is to couple electrical components through algebraic relationships. For example, $I_{bus,1,1} = I_{out,1}$, as the output of the boost converter feeds into voltage bus 1. The second stage is to couple the thermal models by matching appropriate temperatures together: $\mathbb{T}_{a,l,1} = T_{sr,1}$ is an example of this for server 1. This also includes matching appropriate power flows, such as $P_{cool,4} = X_{vcs}COP$. The third stage is to recognize any electro-thermal coupling. The VCS is powered by converter 4, so $P_{vcs} = V_{out,4}I_{out,4}$. To calculate electrical power, $I_{out,4} = V_{out,4}/R_{load,4}$, with $R_{load}$ as a virtual load resistance. Similar definitions exist for converters 5-6 to power the racks, with $P_{s,1} = V_{out,5}I_{out,5}$ and $P_{s,2} = V_{out,6}I_{out,6}$. The total power to the racks is $P_{load} = P_{s,1} + P_{s,2}$. All details for connecting components are provided in Appendix B.

Figures 8 and 9 present the coupled microgrid and data center elements. There are $N_x = 37$ states of the system: $V_{bus,1}$, $V_{bus,2}$, $T_{PV}$, $SOC$, $V_{re,2}$ to $V_{re,4}$, $T_{s,1}$, $T_{s,2}$, $T_{sr,1}$ to $T_{sr,8}$, $T_{ccu,1}$ to $T_{ccu,4}$, $T_{p,1}$ to $T_{p,3}$, $X_{vcs}$, and the twelve $V_{in}$ and $V_{out}$ states of the converters. There are $N_u = 6$ control inputs that are the duty cycles of the converters, $\boldsymbol{u} = [D_1, \ldots, D_6]$. There are $N_d = 2$ exogeneous inputs, $\boldsymbol{d} = [G, T_\infty]$. To further support the format of the model for CCD purposes, the energy dynamics of the system are written as a graph-based model, shown in Equation (20). The format of the model contains a capacitance matrix $\boldsymbol{C}$, incidence matrix $\boldsymbol{\overline{M}}$, and nonlinear power flows $\boldsymbol{P}$ [50,51]. The sink states $\boldsymbol{x}^s = \boldsymbol{d}$.

$$\boldsymbol{C}\dot{\boldsymbol{x}} = -\boldsymbol{\overline{M}}\boldsymbol{P}(\boldsymbol{x}, \boldsymbol{u}, \boldsymbol{x}^s) \quad (20)$$

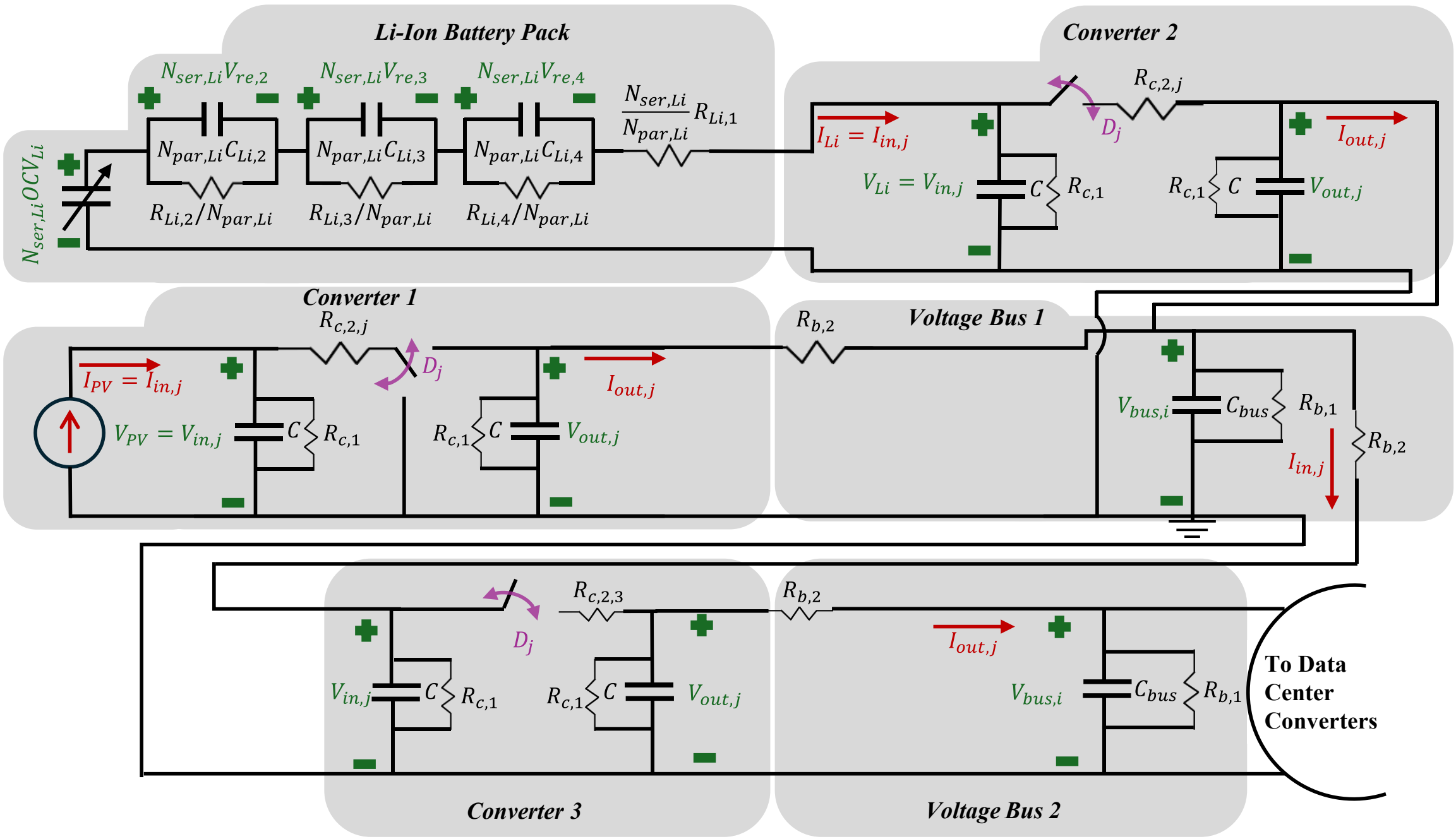

**FIGURE 8:** CIRCUIT DIAGRAM OF MIGROGRID MODEL

## 4. CONTROL ALGORITHM & VALIDATION

This section provides the control algorithm used within the CCD optimization problem. The converter duty cycles are manipulated using a model-free approach. The controller, alongside the microgrid and data center plant model, is tested under realistic simulation conditions.

### 4.1 Model-Free Control Algorithm

Control of the system during operation is accomplished using a model-free strategy. Perturb and observe methods [52] are especially appropriate to compute duty cycle values for power systems. Figure 10 presents the procedure to determine the $i = 1, \dots, N_u$ input values after a timestep $\Delta t$, with time index $k$. Measurements are used to calculate values for $\alpha_i$, the change in the term to be tracked over the timestep, and $\beta_i$, the change in the term that influences $\alpha_i$. Based on $\alpha_i$ and $\beta_i$, the next input values $u_{i,k+1}$ will be one of the following options: it holds the prior value $u_{i,k}$, or it is adjusted from the prior value by $\pm\Delta u$. The constant parameter $\Delta u$ is the duty cycle perturbation. All inputs are saturated to be kept within their bounds, being 0 to 1 for converter duty cycles. There are nine discrete modes for the controller as presented in Appendix C. The mode is determined based on the values of $SOC$, $T_{s,1}$ and $T_{s,2}$, modifying the $\alpha_i$ and $\beta_i$ expressions. Two critical modes are highlighted to provide examples:

(a) In mode 1, the battery $SOC$ is between its lower and upper bounds, $SOC_{min}$ and $SOC_{max}$. Both server rack temperatures are also within lower and upper bounds, $T_{s,min}$ and $T_{s,max}$. The boost converter duty cycle $u_1 = D_1$ is controlled to maximize PV power $P_{PV} = I_{PV}V_{PV}$, with $\alpha_1 = P_{PV,k} - P_{PV,k-1}$, and $\beta_1 = V_{PV,k} - V_{PV,k-1}$. The duty cycle of converters 5 and 6 are used to track server racks electrical power demand, $P_{s,1,ref}$ and $P_{s,2,ref}$ with $P_{s,1} = V_{out,5}I_{out,5}$ and $P_{s,2} = V_{out,6}I_{out,6}$. For converter 5, $\alpha_5 = P_{s,1,ref} - P_{s,1,k}$ and $\beta_5 =$ $u_{5,k} - u_{5,k-1}$, with similar definitions for converter 6. All other duty cycles are held at constant values in this mode.

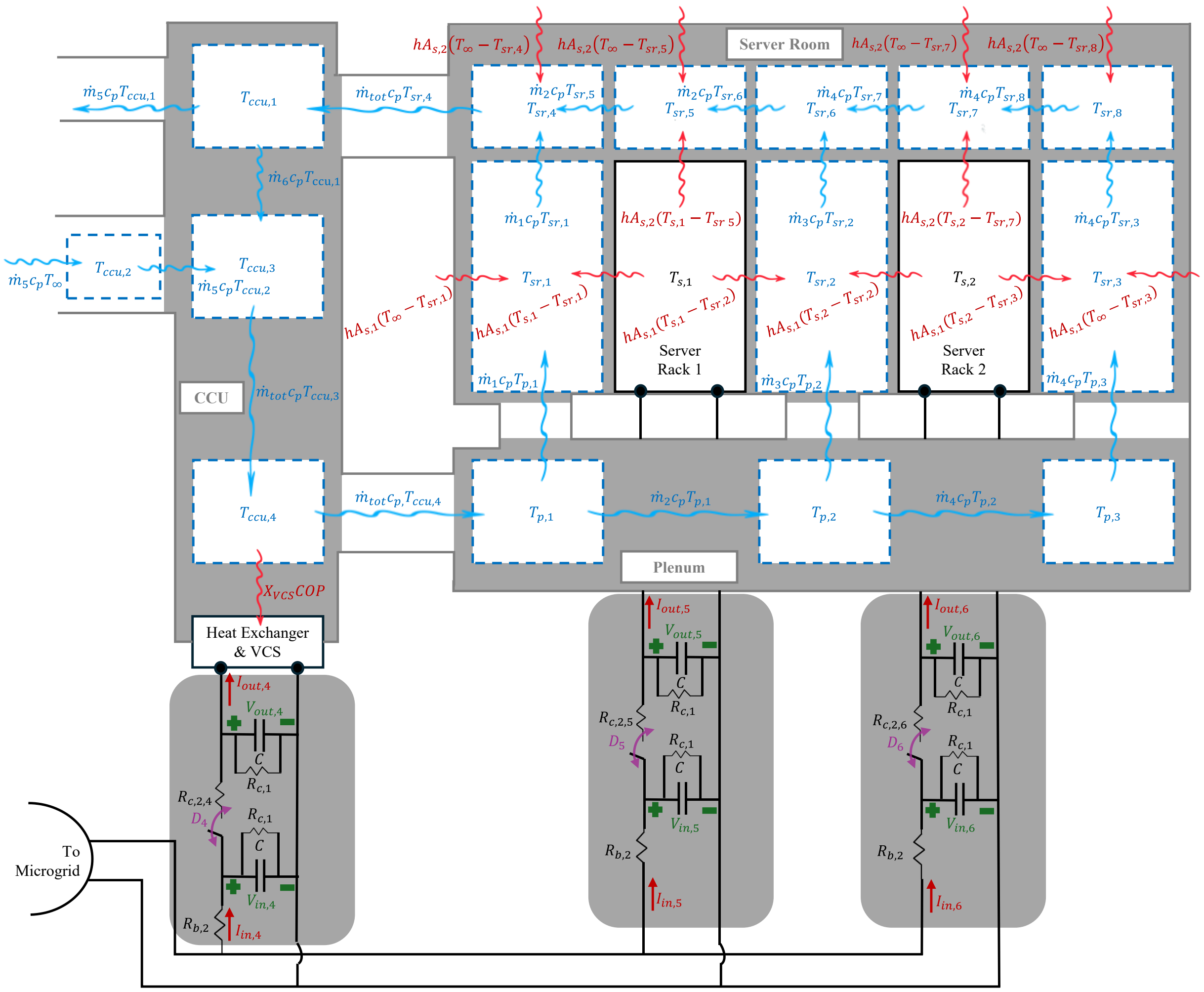

**FIGURE 9:** AIR FLOW DIRECTION INSIDE A DATA CENTER

(b) In mode 9, the battery $SOC$ has exceeded $SOC_{max}$. To prevent overcharging of the battery, the PV is cut off by setting $u_1 = 0$. Converter 2 near the battery is set to track a constant charge reference $SOC_{ref} < SOC_{max}$, with $\alpha_2 = SOC_{ref,k} - SOC_k$ and $\beta_2 = u_{2,k} - u_{2,k-1}$. The VCS converter duty cycle, $u_4 = D_4$, is set to a high constant value to assist in draining the battery to a safe $SOC$. The server rack converters still track $P_{s,1,ref}$ and $P_{s,2,ref}$, and the remaining converter duty cycles are held constant.

Equation (21) presents a representative vector of functions $\boldsymbol{f}$ for this control policy. The vector $\boldsymbol{x}$ contains required states like $SOC$, $T_{s,1}$, and $T_{s,2}$, and the vector $\boldsymbol{P}$ contains power flows like $P_{s,1}$, $P_{s,2}$, and $P_{PV}$ as determined from measurements. References, state bounds, and input bounds are captured in $\boldsymbol{x}_{ref}$, $\boldsymbol{P}_{ref}$, $\boldsymbol{x}_{min}$, $\boldsymbol{x}_{max}$, $\boldsymbol{u}_{min}$, and $\boldsymbol{u}_{max}$.

$$\boldsymbol{u}_{k+1} = \boldsymbol{f}(\boldsymbol{x}_k, \boldsymbol{x}_{k-1}, \boldsymbol{P}_k, \boldsymbol{P}_{k-1}, \Delta u, \boldsymbol{u}_k, \boldsymbol{u}_{k-1}, \boldsymbol{x}_{ref,k}, \boldsymbol{P}_{ref,k}, \boldsymbol{x}_{min}, \boldsymbol{x}_{max}, \boldsymbol{u}_{min}, \boldsymbol{u}_{max}) \tag{21}$$

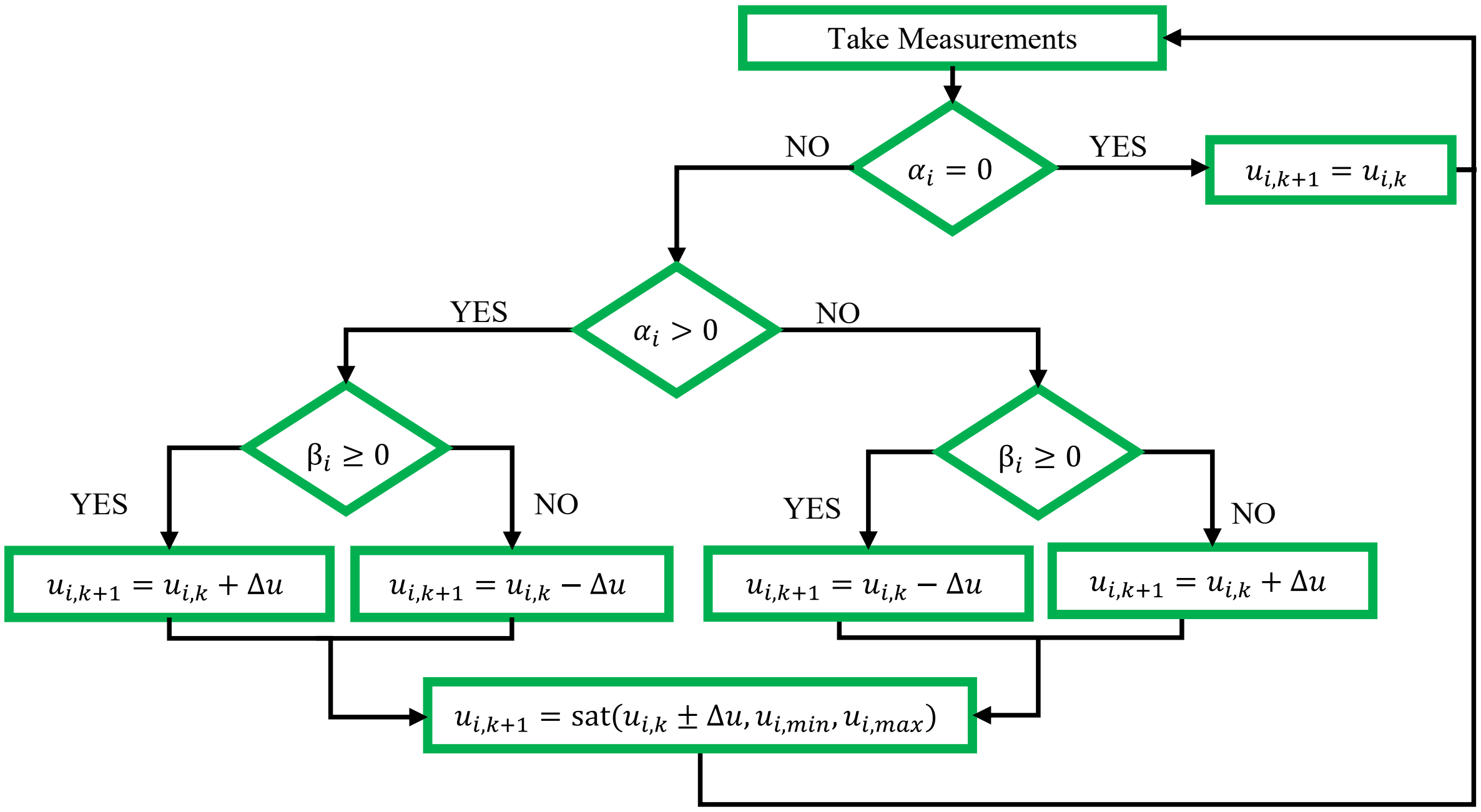

**FIGURE 10:** PERTURB & OBSERVE CONTROL ALGORITHM LOGIC

### 4.2 Controller Validation

The controller's performance is validated by testing the model under small variations in certain parameters. On the plant side, three sets of parameters are used: $N_{par,Li} = 100N_{par,Li,nom}$ and $h_s = h_{s,nom}$, $N_{par,Li} = 400N_{par,Li,nom}$ and $h_s = h_{s,nom}$, and $N_{par,Li} = 400N_{par,Li,nom}$ and $h_s = 1.4h_{s,nom}$. The initial conditions for the battery SOC and server rack temperatures are 0.5 and 18ºC, respectively. The exogeneous inputs are solar irradiance $G$ and ambient temperature $T_\infty$. The irradiance profile is modeled as a sinusoidal function with a peak of 1000 $\text{W/m}^2$ for 12 hours of daylight. The ambient temperature is assumed to be constant in this case. The duty cycle perturbation magnitude $\Delta u$ is a constant 0.01. Additional controller parameters include $\Delta t = 60$ s, $SOC_{min} = 0.1$, $SOC_{max} = 0.9$, $T_{s,min} =$ 13°C, and $T_{s,max} = 80$°C. Reference $SOC_{ref} = 0.5$, and $P_{s,1,ref} = P_{s,2,ref}$ are scaled rack power profiles from [53]. The MATLAB function `ode23tb` is used for simulation, with $t_f = 24$ h.

Figure 11 presents $P_{s,1}$, $SOC$, energy generated by the PV $E_{PV}$, and $T_{s,1}$ during the test. As reflected by the figure, the controller performs as anticipated. All three parameter sets yield designs that track the reference power to the servers for at least 20 hours. Similarly, all three cease charging the battery pack once $SOC_{max}$ is reached. The energy supplied by the PV is curtailed to avoid overcharging the battery. The first parameter set has a lower battery capacity as compared to the other two parameter sets. At 20 hours, the battery reaches $SOC_{min}$, forcing the controller to shut off electrical power to the server racks. This captures the controller's decision to prioritize battery safety over load tracking. The third parameter set has a variation in the number of servers the data center contains. Comparing the profiles for parameter sets 2 and 3 in Figure 11(d) show that the number of servers is inversely proportional to rack temperature. The reason behind this is that the vertical surface area increases, enhancing heat transfer. Additionally, the same amount of input electrical power, correlated with the number of computing jobs scheduled, is distributed to more servers.

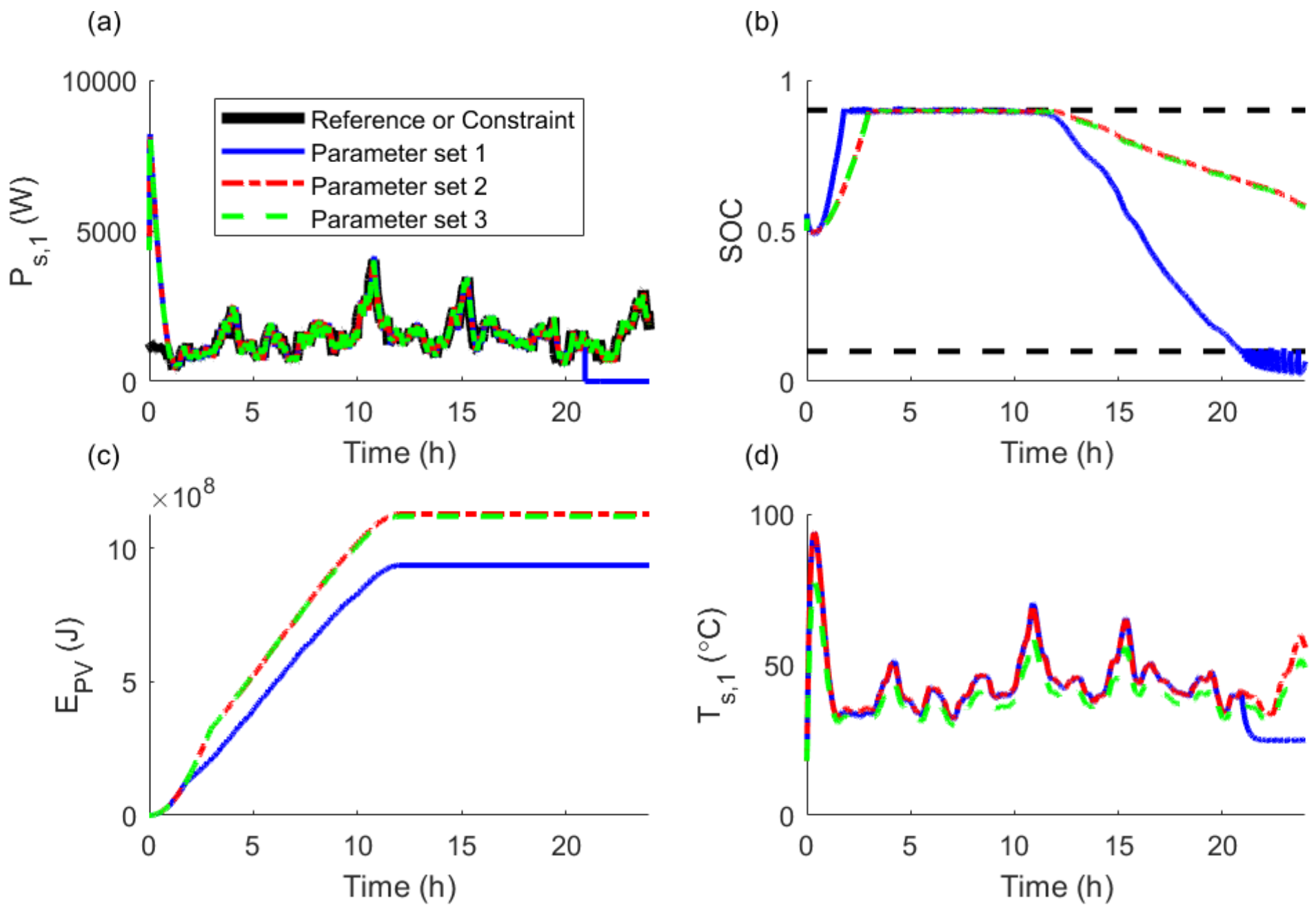

**FIGURE 11:** (a) RACK POWER PROFILE, (b) BATTERY SOC, (c) ENERGY OUT OF THE PV, AND (d) SERVER RACK 1 TEMPERATURE

## 5. APPLICATION OF PROPOSED CCD FRAMEWORK TO THE CANDIDATE SYSTEM

In this section, the proposed sustainability-centric CCD framework is applied to the candidate system to optimize the closed-loop system. Following the first step, $N_\theta = 2$ plant design variables are selected: $\theta_1$ scales the number of Li-ion cells in parallel to have $N_{par,Li} = \theta_1 N_{par,nom}$. The dimensionless scaling parameter $\theta_2$ adjusts the number of servers of both racks $h_s = \theta_2 h_{s,nom}$. This influences surrounding surface areas and volumes of air in server room CVs. This continuous variable captures the number of servers in the data center. The $N_\phi = 1$ controller design variable sets the duty cycle perturbation magnitude $\phi_1 = \Delta u$. This is identical for all the converters of the system.

The second step involves definition and quantification of a family of sustainability objectives, one of the contributions of this paper. Equations (22)-(24) are used to calculate $J_{tot}$ through Equation (1), with $N = 1$ used in this work. *The manufacturing objective function* quantifies GHG emissions during the production of microgrid and data center components. Equation (22) describes the relationship between plant design variables and two terms that make up $J_m$. The function $J_{m,MG}$ calculates emissions from manufacturing battery cells, with $c_{m,MG} = 0.4365 \frac{\text{kgCO}_{2,\text{e}}}{\text{cell}}$ [54]. The function $J_{m,DC}$ is used to determine emissions released from producing servers, with $c_{m,DC} = 9891 \frac{\text{kgCO}_{2,\text{e}}}{\text{rack}}$ [55]. These correlate component sizing with GHG emission equivalents. The units of $J_m$ are kg of $CO_2$ equivalent ($CO_{2,e}$).

$$\begin{aligned} J_m &= J_{m,MG} + J_{m,DC} \\ J_{m,MG} &= c_{m,MG}\theta_1 N_{par,Li,nom} N_{ser,Li} \\ J_{m,DC} &= 2c_{m,DC}\theta_2 \end{aligned} \quad (22)$$

*The operation objective function*, presented in Equation (23), calculates waste heat generated while the plant delivers power to the servers. $P_{out,MG}$ and $P_{out,DC}$ are the total waste heat released at a given time for the microgrid and data center halves, respectively. The coefficients $c_{o,MG} = c_{o,DC} = 3.6 \times 10^{-9} \frac{\text{kgCO}_{2,\text{e}}}{\text{J}}$ convert wasted heat into kg of $\text{CO}_{2,\text{e}}$, as waste heat can have a fraction of the impact of GHG released [56]. $J_o$ is calculated from time $t = 0$ to $t = t_f$.

$$
\begin{aligned}
J_o &= J_{o,MG} + J_{o,DC} \\
J_{o,MG} &= c_{o,MG} \int_0^{t_f} P_{out,MG}(\tau) d\tau \\
J_{o,DC} &= c_{o,DC} \int_0^{t_f} P_{out,DC}(\tau) d\tau
\end{aligned}
\tag{23}
$$

*The disposal objective function* quantifies the amount of e-waste produced each year due to disposing of microgrid and data center components. For the microgrid half, the battery is considered the primary source of e-waste. The objective $J_{d,MG}$ is presented in Equation (24). The numerator of $J_{d,MG}$ describes the mass of the battery pack that will eventually be disposed of, with $m_{cell}$ representing the mass of a Li-ion cell [46]. The denominator of $J_{d,MG}$ predicts the number of years it takes to reach battery end-of-life based on the simulation from 0 to $t_f$. The battery is discarded after losing 20% of its nominal capacity $Q_{Li}$. The capacity loss of a battery cell over the simulation, $Q_{loss}$, is a nonlinear function of $SOC$ and $I_{Li}$ as described in [57,58]. The value $3.154 \times 10^7$ converts seconds into years. Greater battery degradation leads to earlier battery replacement, which in turn generates more e-waste.

For the data center half, failing servers are a source of e-waste. The numerator of $J_{d,DC}$, also presented in Equation (24), contains the mass of the servers, $m_{s,1}$ and $m_{s,2}$, that will be disposed of. $m_{s,1}$ and $m_{s,2}$ are explicitly dependent on increased numbers of servers associated with rack height. These are defined in Appendix A. The denominator is used to determine the number of years until server failure, calculated using the mean time to failure (MTTF) of processor transistors due to thermal cycling. MTTF is itself a function of average server rack temperatures $T_{s,1}$ and $T_{s,2}$, and ambient temperature $T_\infty$ [44,45]. Two parameters are used for MTTF calculation: $C_o$ represents the proportionality constant and $q$ is the Coffin-Manson exponent [44,45]. The objective $J_d$ sums the e-waste from these individual pieces and is in units of kg of e-waste per year.

$$
\begin{aligned}
J_d &= J_{d,MG} + J_{d,DC} \\
J_{d,MG} &= \frac{m_{cell}\theta_1 N_{par,Li,nom} N_{ser,Li}}{0.2Q_{Li}/\left(Q_{loss}(SOC, I_{Li}) \times 3.154 \times 10^7 / t_f\right)} \\
J_{d,DC} &= \frac{m_{s,1}(\theta_2) + m_{s,2}(\theta_2)}{C_0\left(\frac{1}{t_f}\int_0^{t_f} \frac{1}{2}\left(T_{s,1}(t) + T_{s,2}(t)\right) dt - T_\infty\right)^{-q}}
\end{aligned}
\tag{24}
$$

For the third step of the framework, plant dynamics of Equation (20) should first be adjusted to include the influence of plant design variables. Equation (25) presented the augmented model, with $\boldsymbol{\Psi_c(\theta)}$ and $\boldsymbol{\Psi(\theta)}$ as diagonal design matrices [47].

$$
\boldsymbol{\Psi_c(\theta) C \dot{x}} = -\boldsymbol{\bar{M} \Psi(\theta) P(x, u, x^s)}
\tag{25}
$$

Matching terms to Equation (2), $\boldsymbol{E}(\boldsymbol{\theta}) = \boldsymbol{\Psi_c}(\boldsymbol{\theta})\boldsymbol{C}$ and $\boldsymbol{f_P} = -\boldsymbol{\bar{M}}\boldsymbol{\Psi}(\boldsymbol{\theta})\boldsymbol{P}(\boldsymbol{x}, \boldsymbol{u}, \boldsymbol{x^s})$, again with $\boldsymbol{x^s} = \boldsymbol{d}$. The constraint for the initial conditions retains the form of Equation (3), and the controller constraint is defined by matching Equations (4) and (21). The key inequality constraint for this optimization problem is to assure that power tracking for the servers is satisfied. Traditionally, $\mu_1 = J_{tr}$ as defined in Equation (26) is an objective function to minimize. With sustainability as the focus, instead it is desired to keep this under $\mu_{1,max} = 5.3 \times 10^{10}$, with this value being selected based on the tests in Section 4.

$$\mu_1 = J_{tr} = \frac{1}{2}\sum_{i=1}^{2}\int_0^{t_f}\left(P_{s,i,ref}(\tau) - P_{s,i}(\tau)\right)^2 d\tau \tag{26}$$

The remaining inequality constraints signify the critical limit in the plant and controller design variables. Equations (27)-(28) indicate these bounds.

$$\begin{gathered} \boldsymbol{\theta_{min}} = [100, 0.5] \\ \boldsymbol{\theta_{max}} = [2200, 1.5] \\ \boldsymbol{\theta_{min}} \leq \boldsymbol{\theta} \leq \boldsymbol{\theta_{max}} \end{gathered} \tag{27}$$

$$\begin{gathered} \boldsymbol{\phi_{min}} = [0.01] \\ \boldsymbol{\phi_{max}} = [0.03] \\ \boldsymbol{\phi_{min}} \leq \boldsymbol{\phi} \leq \boldsymbol{\phi_{max}} \end{gathered} \tag{28}$$

Based on the factors defined above, the fourth step organizes the CCD problem into a proper structure, shown in Equation (29). Note that $k$ is the time index related to $\Delta t$ and $t$.

$$\begin{aligned} \min_{\boldsymbol{\theta}, \boldsymbol{\phi}} \quad & J_{tot} = w_m J_m + w_o J_o + w_d J_d \\ \text{s.t.} \quad & \boldsymbol{\Psi_c}(\boldsymbol{\theta})\boldsymbol{C}\dot{\boldsymbol{x}} + \boldsymbol{\bar{M}}\boldsymbol{\Psi}(\boldsymbol{\theta})\boldsymbol{P}(\boldsymbol{x}, \boldsymbol{u}, \boldsymbol{x^s}) = \boldsymbol{0}, \forall\, t \in [0, t_f] \\ & \boldsymbol{u}_{k+1} = \boldsymbol{f}(\boldsymbol{x}_k, \boldsymbol{x}_{k-1}, \boldsymbol{P}_k, \boldsymbol{P}_{k-1}, \Delta u, \boldsymbol{u}_k, \boldsymbol{u}_{k-1}, \ldots \\ & \qquad \boldsymbol{x_{ref,k}}, \boldsymbol{P_{ref,k}}, \boldsymbol{x_{min}}, \boldsymbol{x_{max}}, \boldsymbol{u_{min}}, \boldsymbol{u_{max}}), \\ & \qquad \forall\, k \in [0, t_f/\Delta t] \\ & \boldsymbol{x}(0) - \boldsymbol{x_{IC}} = \boldsymbol{0} \\ & \mu_1 \leq \mu_{1,max} \\ & \boldsymbol{\theta}_{min} \leq \boldsymbol{\theta} \leq \boldsymbol{\theta}_{max} \\ & \boldsymbol{\phi}_{min} \leq \boldsymbol{\phi} \leq \boldsymbol{\phi}_{max} \end{aligned} \tag{29}$$

In the fifth step, this work uses a grid search to optimize the design variables. A selected number of values are permitted for each variable between the respective minimum and maximum values: $\theta_1$ in increments of 100, $\theta_2$ in increments of 0.05, and $\phi_1$ in increments of 0.01. $\theta_2$ is held constant while $\theta_1$ is varied, and vice-versa; $\phi_1$ is always varied. This makes for 129 designs. Additional 37 randomized combinations and 104 Latin hypercube sampling of the design variable values are also tested, making for a total of 270 designs. The shooting method computes the objective by simulating the plant and controller dynamics for each set of design variable values. The optimization procedure involves checking each design to ensure constraints are satisfied. The other part of the fifth step provides the analysis of the outcomes of this CCD problem, which is discussed in the next section.

## 6. RESULTS AND DISCUSSION

This section presents the outcomes of implementing the CCD framework on the candidate system. From searching the design space, potential solutions are assessed to develop a relationship between the design variables and the sustainability objectives. To identify trade-offs among the objectives, 2D Pareto fronts are constructed from the Pareto-optimal solutions. Comparison is made in part by determining the sensitivity of objective functions with respect to each other along the front. One design is compared against a design identified from a baseline procedure.

### 6.1 Relationships Between Design Variables and Objective Functions

The grid search permits exploration of behavioral trends between the design variables and sustainability objective functions. Three designs with $\theta_1 < 200$ exceed the power tracking error constraint of $\mu_{1,max}$ and are considered unacceptable. The following observations are made by analyzing the remaining designs that vary $\theta_1$ and $\phi_1$:

(a) Figure 12 shows that the manufacturing objective function relies solely on the plant design variable, as expected from Equation (22).

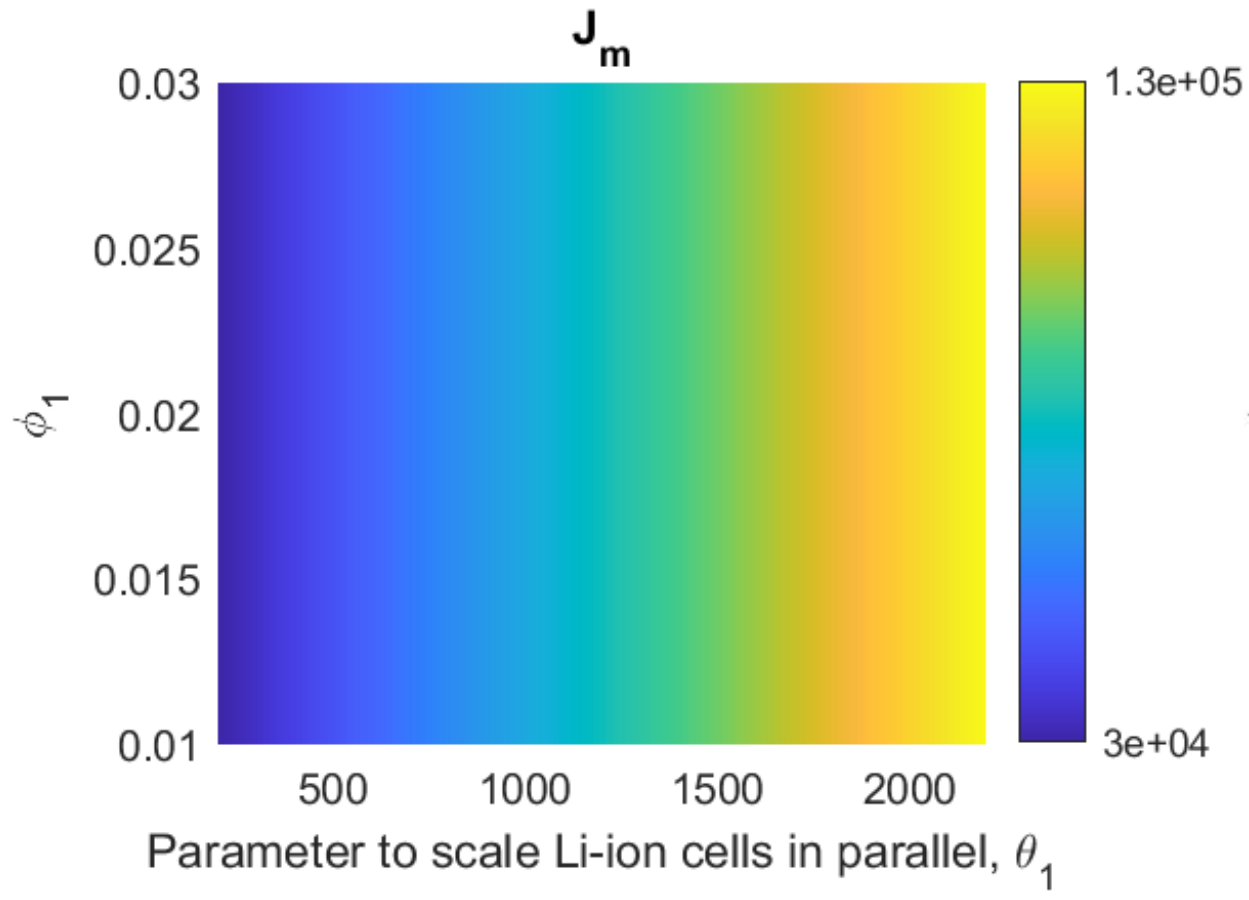

**FIGURE 12:** $J_m$ VERSUS DESIGN VARIABLES $\theta_1$, $\phi_1$

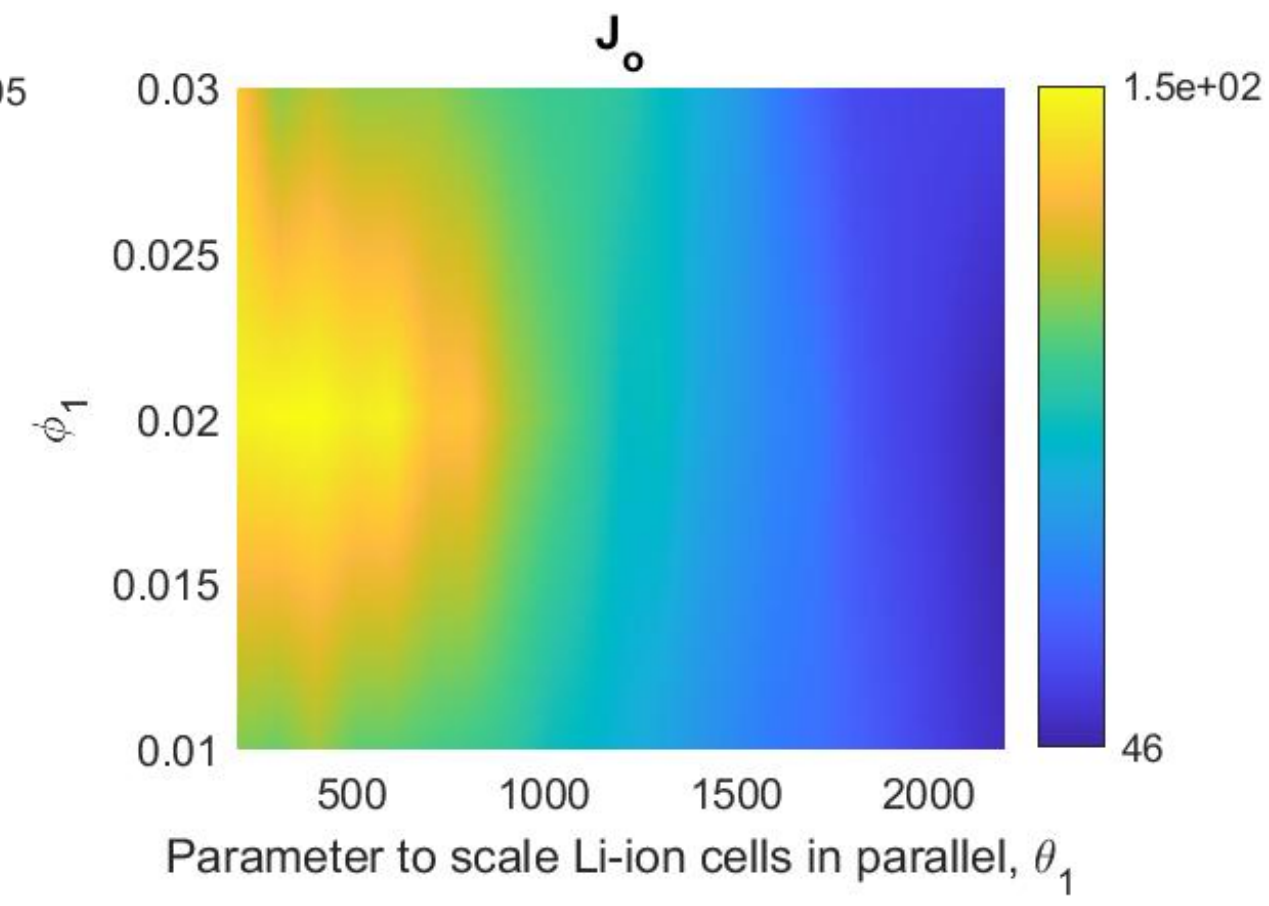

**FIGURE 13:** $J_o$ VERSUS DESIGN VARIABLES $\theta_1$, $\phi_1$

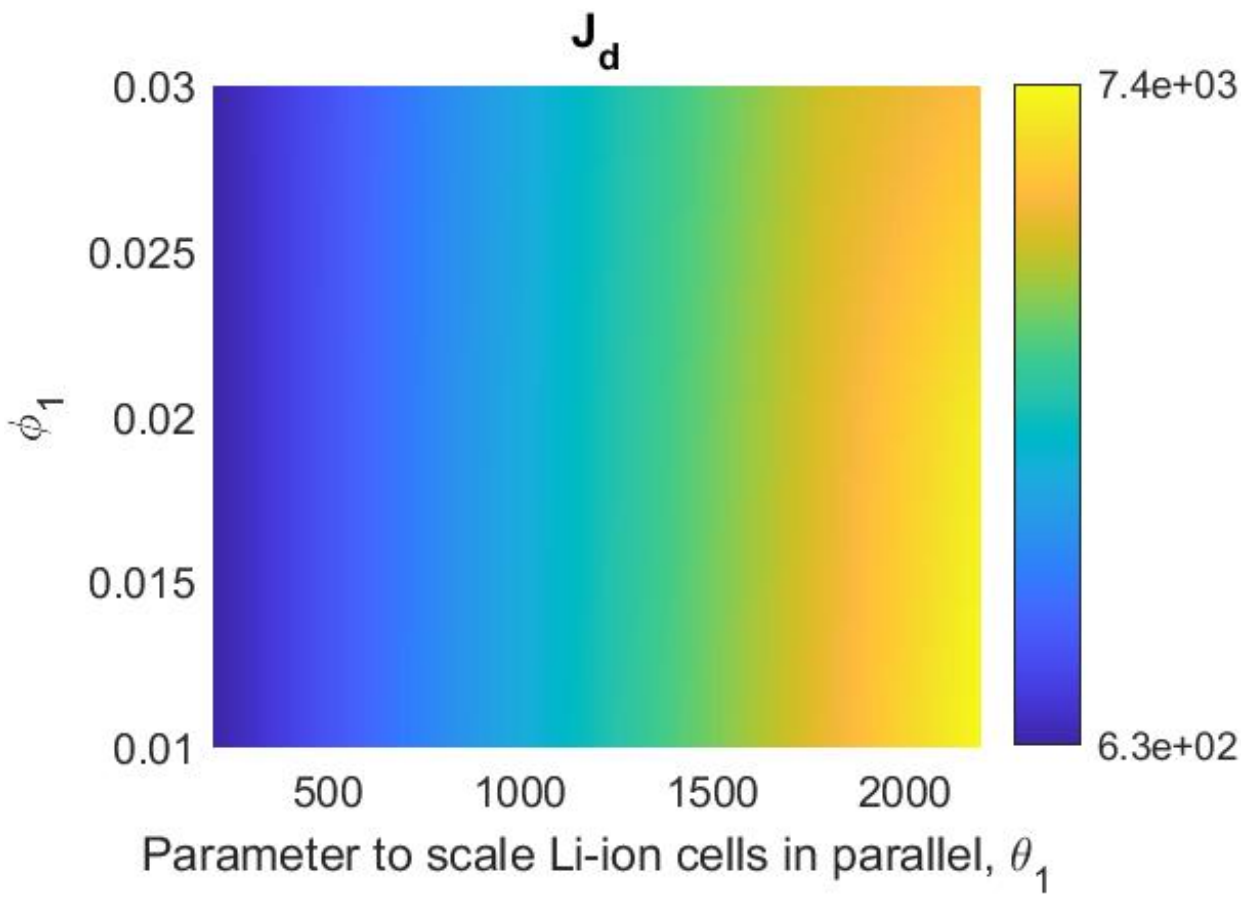

**FIGURE 14:** $J_d$ VERSUS DESIGN VARIABLES $\theta_1$, $\phi_1$

(b) Figure 13 shows that the operation objective is influenced by both $\theta_1$ and $\phi_1$. The sensitivity is checked at $\phi_1 = 0.02$ and two distinct values of $\theta_1$. A value of $\frac{\partial J_o}{\partial \theta_1} = -0.01345$ is found for $\theta_1 = 500$, and $\frac{\partial J_o}{\partial \theta_1} = -0.08765$ for $\theta_1 = 2000$. This indicates that changes in the plant design variable are more impactful on waste heat generation as $\theta_1$ increases. Calculated in a similar manner, $\frac{\partial J_o}{\partial \phi_1} = 284.5$ for $\theta_1 = 500$ and $\frac{\partial J_o}{\partial \phi_1} = 16.50$ for $\theta_1 = 2000$. This indicates that the operation objective is sensitive to changes in the controller design variable while the plant design variable value is low.

(c) Figure 14 shows that the disposal objective is greatly impacted by $\theta_1$. The trend is similar to that shown for the manufacturing objective in Figure 12. The difference is that the controller design variable has a slight influence on $J_d$ at higher values of the plant design variable.

The second set of observations is drawn from varying $\theta_2$ and $\phi_1$:

(a) Also matching with Equation (22), Figure 15 shows that the controller design variable does not influence the manufacturing objective when adjusting $\theta_2$.

(b) Figure 16 shows that adjusting the number of servers has minimal influence on the operation objective of waste heat except at high $\phi_1$ values. The sensitivity $\partial J_o/\partial \theta_2$ reaches a peak of 158.0 at approximately $\theta_2 = 0.75$ and $\phi_1 = 0.03$. The controller design variable $\phi_1$ more heavily influences $J_o$ as compared to $\theta_2$.

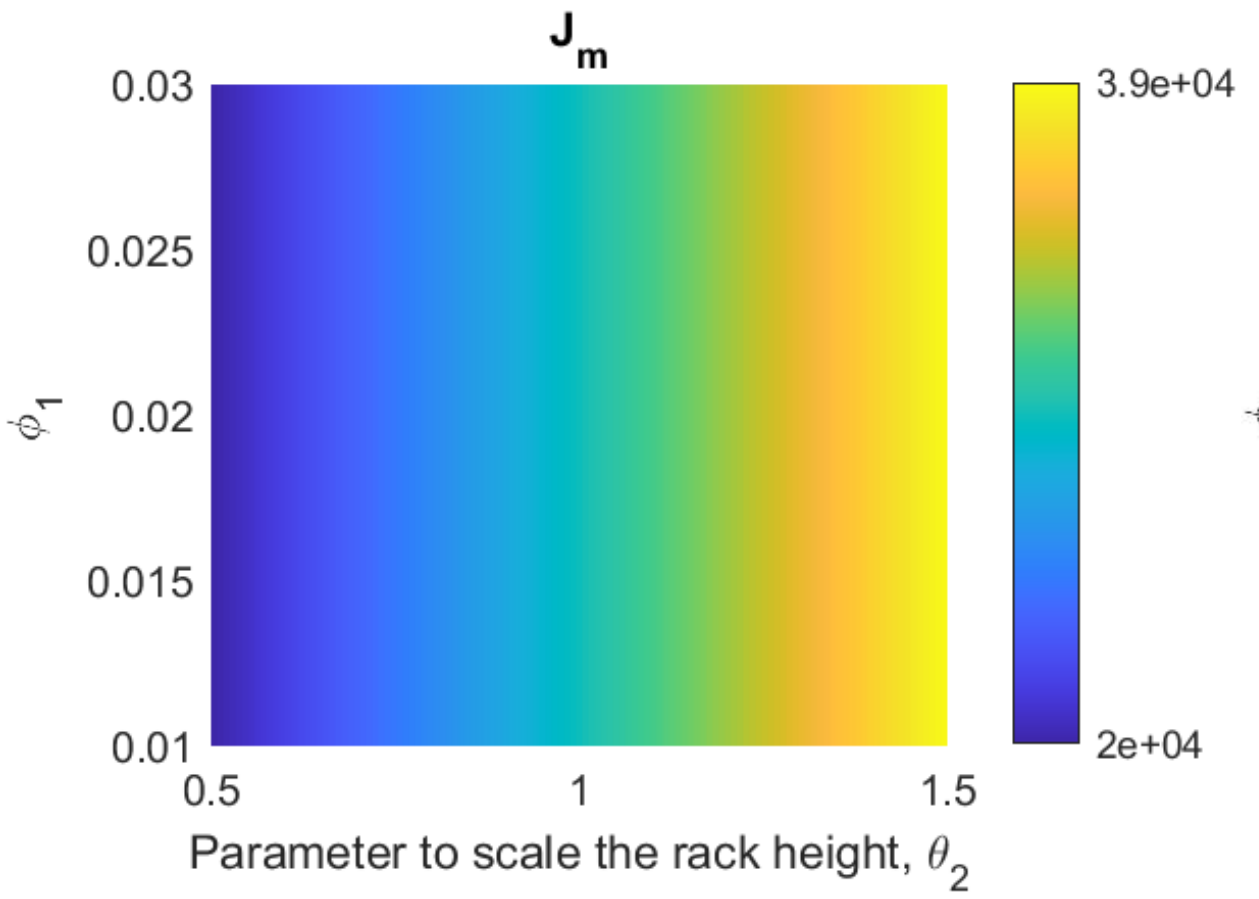

**FIGURE 15:** $J_m$ VERSUS DESIGN VARIABLES $\theta_2$, $\phi_1$

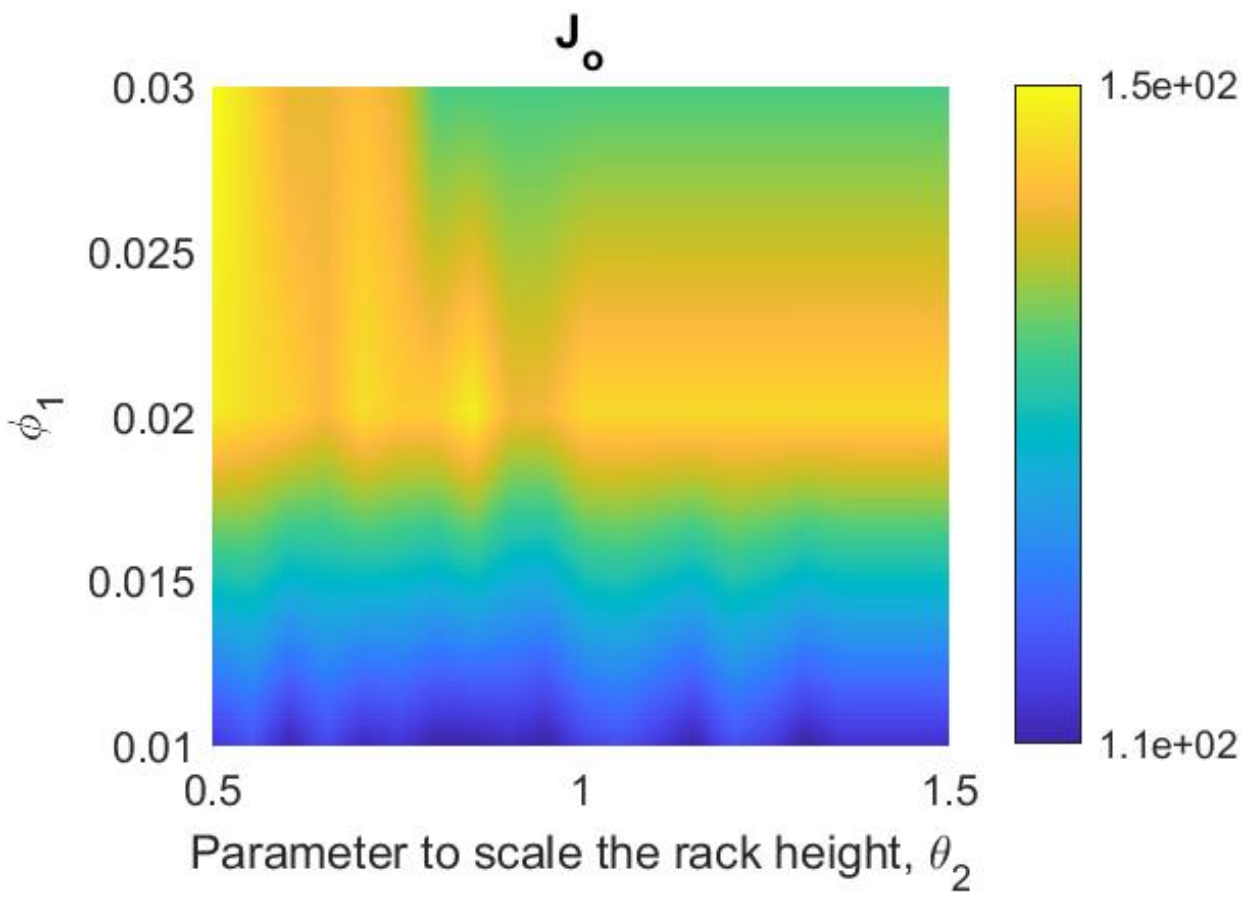

**FIGURE 16:** $J_o$ VERSUS DESIGN VARIABLES $\theta_2$, $\phi_1$

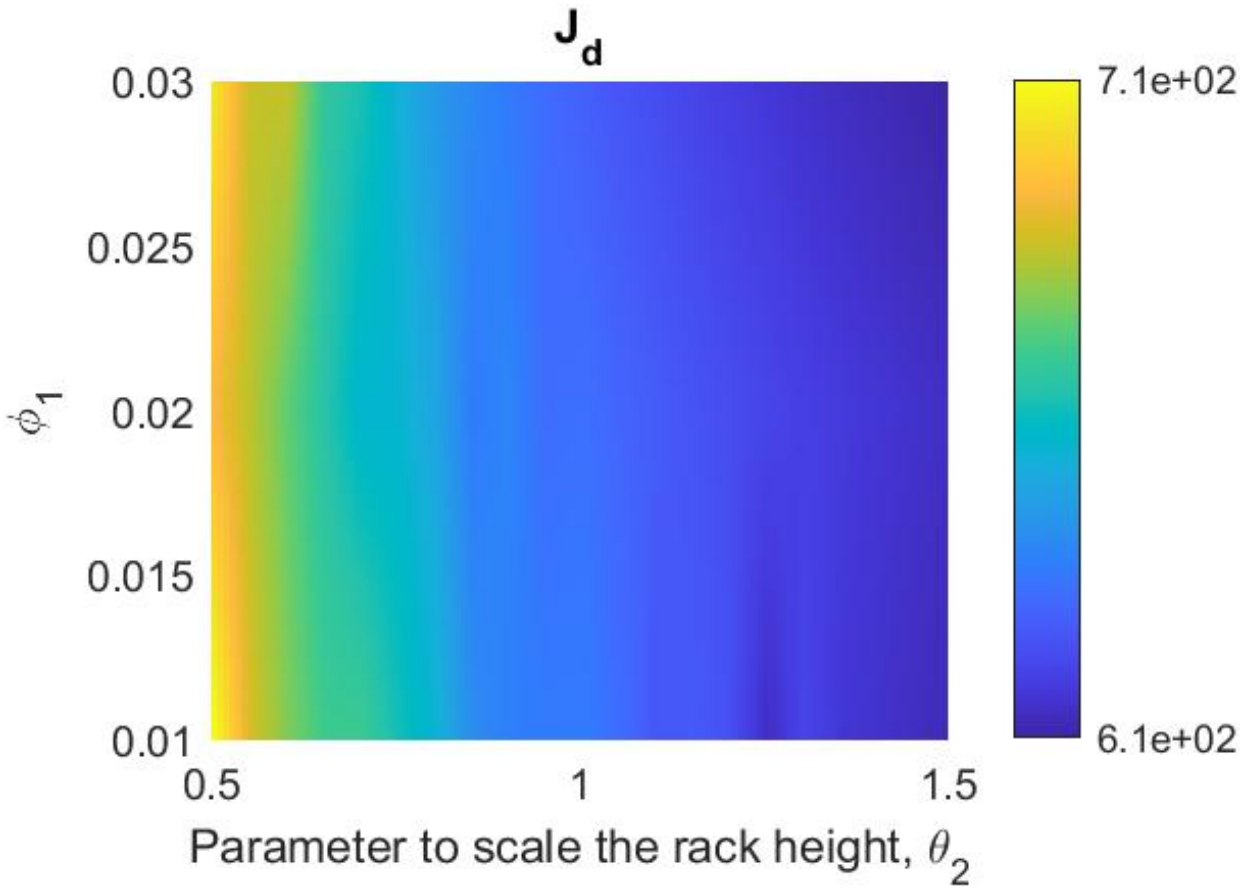

**FIGURE 17:** $J_d$ VERSUS DESIGN VARIABLES $\theta_2$, $\phi_1$

(c) Figure 17 shows that generated e-waste is more heavily influenced by $\theta_2$ as compared to $\phi_1$ for this subset of designs. However, as compared to changing the battery pack size, the trend is the opposite: increasing the number of servers ultimately reduces the disposal objective.

Studying Figures 12-14 with their equivalents in Figures 15-17 allows comparison of the microgrid plant design variable, $\theta_1$, and the data center plant design variable, $\theta_2$. Manufacturing is more sensitive to $\theta_2$ as compared to $\theta_1$, as $c_{m,DC} > c_{m,MG}$. However, $\theta_1$ has a larger feasible range of values, giving it more influence over the objective's value. For the operation objective, Figures 13 and 16 show a notable trend: changing $\phi_1$ does not impact $J_o$ when changing $\theta_1$. However, changing $\phi_1$ can have an influence on $J_o$ depending on the value of $\theta_2$. Comparing Figures 14 and 17, $\theta_1$ and $\theta_2$ have contrary impacts on $J_d$. Increasing the number of battery cells increases degradation as smaller battery packs hit lower SOC limits earlier and shut down the system. Increasing the number of servers reduces the workload for each server, driving its average temperature and thus degradation down. Higher degradation leads to more e-waste or early component replacement, which is detrimental.

### 6.2 Sustainability-Centric Pareto Fronts

This part of the analysis compares competing sustainability objective functions against each other. Figure 18 compares $J_m$ vs. $J_o$ for all designs, the same as setting $w_d = 0$. The average sensitivity across the Pareto front is $\partial J_m / \partial J_o = -8.217 \times 10^2$. These objectives are both in terms of kg of $CO_{2,e}$, with $J_m$ being for a one-time, beginning-of-life manufacturing process, and $J_o$ for a single day of operation. This similarity in units permits a direct comparison of designs along the Pareto front. Consider the two most extreme designs along Figure 18: the leftmost has $J_m = 1.189 \times 10^5$ and $J_o = 39.04$; the rightmost has $J_m = 1.987 \times 10^4$ and $J_o = 159.6$. The $J_m$ and $J_o$ values for each design can be summed by extrapolating the waste heat values across the expected lifespan of the components, yielding a total kg of $CO_{2,e}$ per design. If the number of years of operation without a need to manufacture replacement components is 2.25 years, these two designs have equal impact on the environment. If components need to be replaced in under 2.25 years, the rightmost design should be selected to prioritize reduction of manufacturing emissions. If components last longer than 2.25 years, the leftmost design should be selected, prioritizing minimization of waste heat emissions.

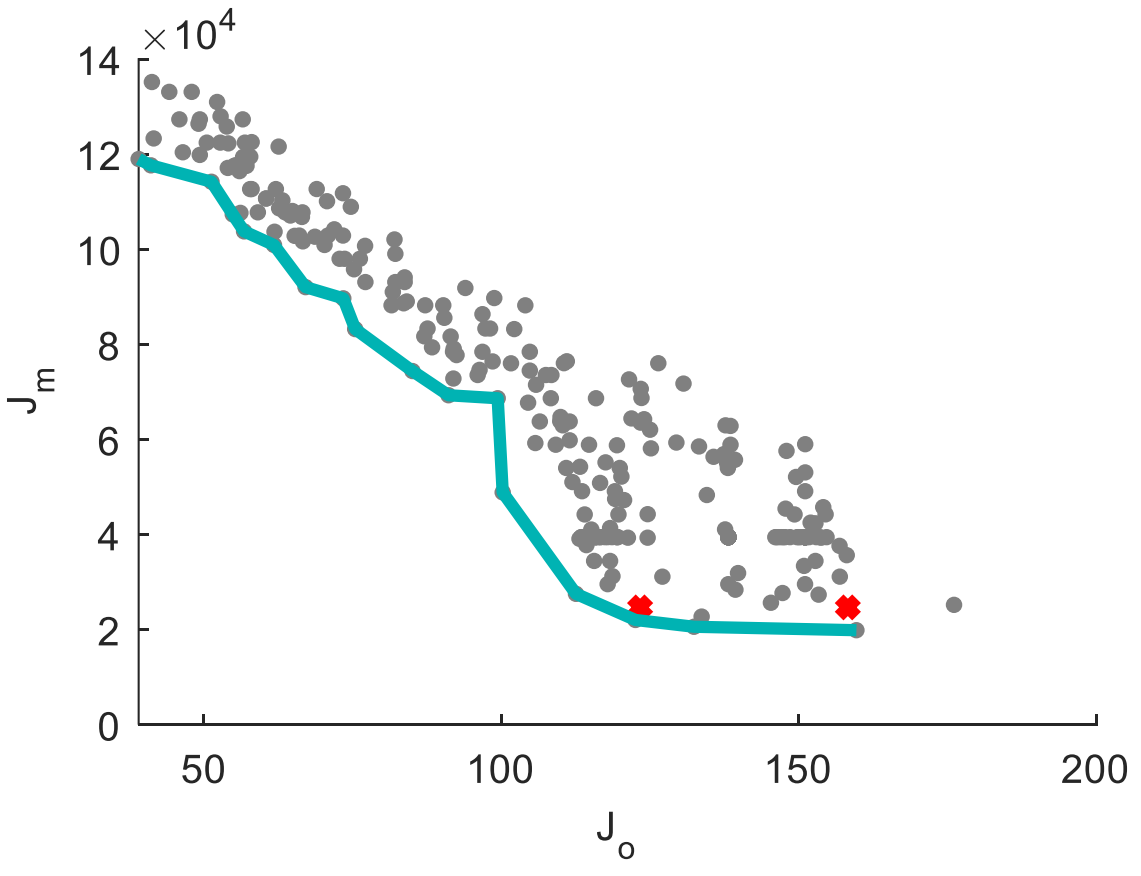

**FIGURE 18:** DESIGN OPTIONS (GREY DOTS), PARETO FRONT (TEAL LINE) FOR $w_d = 0$, AND FAILED DESIGNS (RED X-MARK)

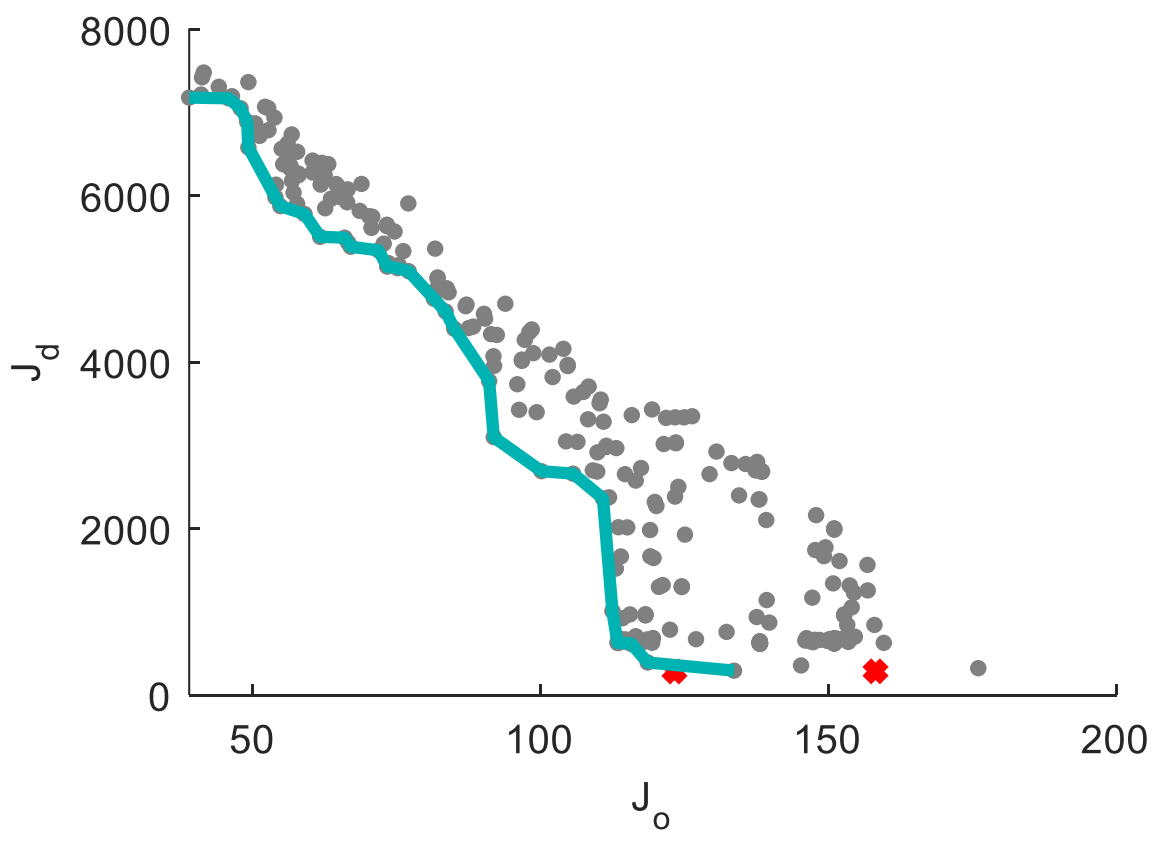

**FIGURE 19:** DESIGN OPTIONS (GREY DOTS), PARETO FRONT (TEAL LINE) FOR $w_m = 0$, AND FAILED DESIGNS (RED X-MARK)

Figure 19 compares $J_d$ vs. $J_o$ for all designs, with $w_m = 0$. The sensitivity across the Pareto front is $\frac{\partial J_d}{\partial J_o} = -72.81$, but dissimilarity in objective function units requires additional steps for comparison of designs. $J_d$ is determined based on the amount of e-waste produced in one year with the unit being kg of e-waste, and $J_o$ is in kg of $CO_{2,e}$ per day as stated earlier. Selection of weights $w_o$ and $w_d$ should be informed by a conversion rate $c_{d,o}$ that represents a ratio of kg $CO_{2,e}$ over kg e-waste. To determine the threshold value of $c_{d,o}$, the two extreme designs along Figure 19 are assumed to have equal environmental impact (i.e., equal $J_{tot}$ values). With $w_m = 0$, $w_o = 1$, and $w_d = \frac{c_{d,o}}{365.25}$ to convert into kg per day, and $N = 1$, Equation (1) is calculated twice: for the leftmost design ($J_d = 7.181 \times 10^3$ and $J_o = 39.04$) and the rightmost design ($J_d = 2.927 \times 10^2$ and $J_o = 133.6$). These two expressions are set equal to each other, permitting the determination of $c_{d,o} = 5.02$ as a threshold value. This says that at $c_{d,o} = 5.02$, both the designs have the same environmental impact. For a design engineer, if they're given a value of $c_{d,o} < 5.02$ from an environmental scientist, the leftmost design is preferred as waste heat impacts the environment more than e-waste. However, when $c_{d,o} > 5.02$, the engineer should select the rightmost design to prioritize e-waste reduction. This threshold value of $c_{d,o}$ provides guidance for engineers as they attempt to balance waste heat or GHG emissions against e-waste based on available e-waste vs. emissions data.

### 6.3 Comparison of a Baseline and the Proposed Method

To evaluate sustainability-centric CCD against standard approaches, a baseline optimization problem is solved. In baseline approaches, power tracking is the primary performance criteria, as it relates to running and completing computing jobs assigned to the data center. Rather than a constraint, $J_{tot} = J_{tr}$ is minimized, and sustainability is not a criterion of evaluation. Table 1 compares the optimal baseline design against one Pareto-optimal design produced by solving the proposed, sustainability-centric problem formulation. The baseline design tracks power marginally better – the proposed has a 0.1055% increase. However, both are close to the desired such that job completion would be nearly equivalent. A significant change is observed in the manufacturing objective value, with the proposed having a 38% reduction over the baseline. Figure 20 shows why this is substantial: by prioritizing $J_{tr}$, the baseline gets the design the furthest to the left on the $J_m$ vs. $J_{tr}$ Pareto front. By sacrificing a little power tracking and choosing a more right-side Pareto point, GHG emissions drop by almost an order of magnitude. Furthermore, due to the reduced battery pack size, the amount of kg of e-waste produced in one year is reduced by approximately 22% from selecting the proposed design. The operation objective is reduced by approximately 10% as well. These enhancements underline the significance of including sustainability attributes in the design process.

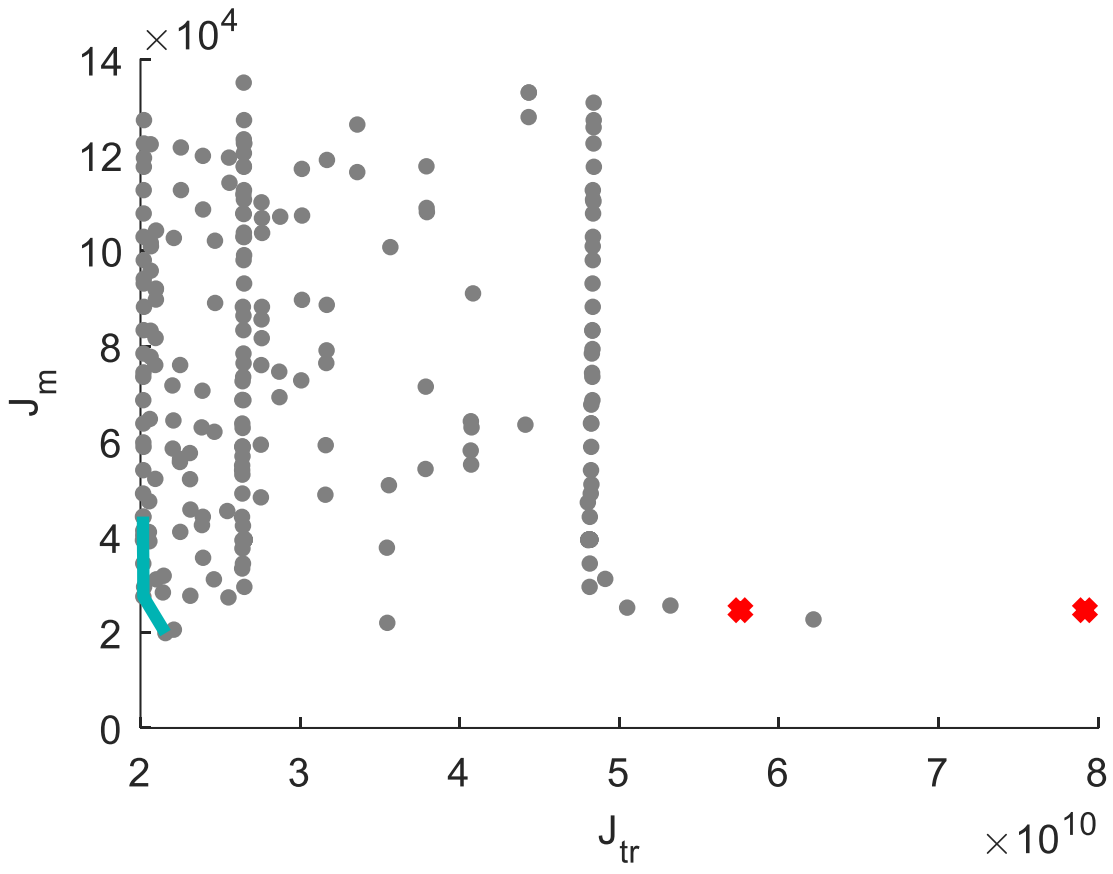

**FIGURE 20:** DESIGN OPTIONS (GREY DOTS), PARETO FRONT (TEAL LINE) AND FAILED DESIGNS (RED X-MARK)

**Table 1:** REDUCTIONS IN OBJECTIVE FUNCTION VALUES AND SIZE OF COMPONENTS AS COMPARED TO THE BASELINE

| | Baseline | Proposed | % Reduction or Improvement |
|---|---|---|---|
| $\theta_1$ | 400 | 300 | 25% reduction |
| $\theta_2$ | 1.25 | 0.65 | 48% reduction |
| $\phi_1$ | 0.03 | 0.03 | 0% reduction |
| $J_m$ | $4.428 \times 10^4$ | $2.752 \times 10^4$ | 37.84% reduction |
| $J_o$ | 124.6 | 112.5 | 9.669% reduction |
| $J_d$ | $1.297 \times 10^3$ | $1.009 \times 10^3$ | 22.17% reduction |
| $J_{tr}$ | $2.016 \times 10^{10}$ | $2.018 \times 10^{10}$ | 0.1055% increase |

## 7. CONCLUSION

This work explores the use of CCD in microgrids and data centers to improve sustainability. While the literature contains studies on plant and controller optimization for these energy systems to increase efficiency, the impacts on sustainability are overlooked. As the lifecycle stages of components contribute to environmental emissions, this work extends the development of a CCD framework for a microgrid-driven data center, with a focus on sustainability. The developed CCD-oriented model is connected to a perturb-and-observe algorithm to control the system dynamics. The CCD problem is solved to optimize plant design variables, including battery pack size, the number of servers, and controller design variables such as duty-cycle perturbation. Three objective functions categories, manufacturing, operation, and disposal, make up the total objective for the proposed framework. The results indicate which designs can be selected to reduce emissions by comparing sustainability objectives against one another for the system analyzed. Compared to baseline designs that minimize power tracking error, the CCD framework can yield up to a 38% reduction in emissions. Future work will explore hardware-in-the-loop (HIL) for validation and comparison of design options.

## ACKNOWLEDGEMENTS

This material is based upon work supported by the National Science Foundation under Award No. 2324707.

## APPENDIX

### A. Model Parameters and Maps

This section presents the maps used in the battery cell and reference power profile. Table A.1 presents parameters of the microgrid, data center model simulation, and experimental data [12,41–43]. Figure A.1 presents the open-circuit voltage, $OCV_{Li}$, as a function of SOC [43]. Figure A.2 represents a scaled rack power demand profile for $P_{s,1,ref} = P_{s,2,ref}$ [53]. Figure A.3 represents the relationship between $COP$ and the load temperature.

The server rack serves as a single control volume with lumped thermal capacitance $C_s$. Thermal capacitance of each server is approximated using a mixture of air (subscript $a$) and aluminum (subscript $al$), with $c_p$ as the specific heat and $m$ as the mass to yield $C_s = 0.98 m_a c_{p,a} + 0.02 m_{al} c_{p,al}$. Mass of each server is $m_{s,i} = m_a + m_{al}$, where $m_a = \rho_a \theta_2 h_{s,nom} t_s l_s$ and $m_{al} = \rho_{al} \theta_2 h_{s,nom} t_s l_s$ with $\rho$ as density and $l_s$ as server length. Each server room air control volume has a thermal capacitance $C_{sr} = m_{a,1} c_{p,a}$ for the $j = [1..3]$ and $C_{sr} = m_{a,2} c_{p,a}$ for the $j = [4..8]$. Similarly, the thermal capacitances for the CCU and plenum air are $C_{CCU} = m_{ccu} c_{p,a}$ and $C_p = m_p c_{p,a}$.

### B. Model Coupling

This section presents the coupling terms that connect the components in the microgrid and data center model. Defining the currents for the two buses as relating to the converter currents, we have: $I_{bus,1,1} = I_{out,1}$, $I_{bus,2,1} = I_{out,2}$, $I_{bus,3,1} = -I_{in,3}$, $I_{bus,4,1} = 0$, $I_{bus,1,2} = I_{out,3}$, $I_{bus,2,2} = -I_{in,4}$, $I_{bus,3,2} = -I_{in,5}$, $I_{bus,4,2} = -I_{in,6}$. Note that the negative terms mean the current is being drawn from the bus.

**Table A.1:** PARAMETERS OF THE SYSTEM, DERIVED FROM (a) [41,42], (b) [43], (c) [46], (d) [44,45], AND (e) [12]

| Symbol | Parameter | Value |
|---|---|---|
| | **Voltage bus** | |
| $C_{bus}$ | Capacitance of the voltage bus | 1000 F |
| $R_{b,1}$ | Resistance | $10^4$ Ω |
| $R_{b,2}$ | Resistance | $10^{-4}$ Ω |
| | **DC/DC converter** | |
| $C$ | Capacitance | 1000 F |
| $R_{c,1}$ | Resistance | $10^4$ Ω |
| $R_{c,2,1}$ | Resistance | $5 \times 10^{-5}$ Ω |
| $R_{c,2,2}$ | Resistance | $5 \times 10^{-4}$ Ω |
| | **PV module** | |
| $N_{par,PV}$ | No. of solar panels in parallel | 500 |
| $N_{ser,PV}$ | No. of solar panels in series | 54 |
| $C_{PV}$ (a) | Lumped thermal capacitance | 4580 J/K |
| $\alpha_{PV}$ (a) | Absorptivity | 0.7 |
| $A_{PV}$ (a) | Module surface area | 0.8 $m^2$ |
| $h_{PV}$ (a) | Convection coefficient | 13.4 W/($m^2$K) |
| $T_\infty$ | Ambient Temperature | 25°C |
| | **Battery** | |
| $N_{ser,Li}$ | No. of Li-ion cell in series | 112 |
| $N_{par,Li,nom}$ | No. of Li-ion cell in parallel | 1 |
| $Q_{Li}$ (b) | Nominal capacity of a cell | 4356 C |
| $C_{Li,2}$ (b) | Capacitance | 15000 F |
| $C_{Li,3}$ (b) | Capacitance | 8000 F |
| $C_{Li,4}$ (b) | Capacitance | 550 F |
| $R_{Li,1}$ (b) | Resistance | 0.0192 Ω |
| $R_{Li,2}$ (b) | Resistance | 0.219 Ω |
| $R_{Li,3}$ (b) | Resistance | 0.0257 Ω |
| $R_{Li,4}$ (b) | Resistance | 0.0331 Ω |
| $m_{cell}$ (c) | Mass | 0.0421 kg |
| | **Server rack** | |
| $q$ (d) | Coffin-Manson exponent | 2.35 |
| $C_o$ | Proportionality constant | $3.44 \times 10^3 \frac{\text{years}}{°\text{C}^{-2.35}}$ |
| $A_{s,1}$ | Side surface area | 1.79 $m^2$ |
| $A_{s,2}$ | Ceiling surface area | 0.56 $m^2$ |
| $h_{s,nom}$ | Nominal height | 1.96 m |
| $t_s$ | Thickness | 0.9 m |
| $l_s$ | Length | 0.61 m |
| $\rho_a$ | Density of air | 1.293 kg/$m^3$ |
| $m_a$ | Mass of air for $i = [1,2]$ | 59 kg |
| $c_{p,a}$ | Specific heat of air | $1004 \frac{\text{J}}{\text{kgK}}$ |
| $\rho_{al}$ | Density of aluminum | 2710 kg/$m^3$ |
| $m_{al}$ | Mass of aluminum | 1.35 kg |
| $c_{p,al}$ | Specific heat of aluminum | $900 \frac{\text{J}}{\text{kgK}}$ |
| $R_{load,i}$ | Virtual load resistance for $i = [1,2]$ | 0.44 Ω |
| $h$ | Convection coefficient | $13.4 \frac{\text{W}}{\text{m}^2\text{K}}$ |
| $\alpha$ | Fraction of air expelled | 1 |
| $\beta$ | Airflow fraction through plenum | 0.33 |
| $\gamma$ | Plenum airflow split fraction | 0.5 |
| $\dot{m}_{tot}$ (e) | Mass flow rate of air | 0.43 kg/s |
| | **Server room air** | |
| $A_{sr,1}$ | Side surface area | 1.79 $m^2$ |
| $A_{sr,2}$ | Ceiling surface area | 0.56 $m^2$ |
| $m_{a,1}$ | Mass of air for $j = [1..3]$ | 1.41 kg |
| $m_{a,2}$ | Mass of air for $j = [4..8]$ | 0.84 kg |
| | **CCU** | |
| $m_{ccu}$ | Mass of air for $k = [1..4]$ | 0.84 kg |
| | **Plenum** | |
| $m_p$ | Mass of air for $n = [1..3]$ | 0.44 kg |
| | **VCS** | |
| $R_{load,4}$ | Virtual load resistance of the VCS | 0.25 Ω |

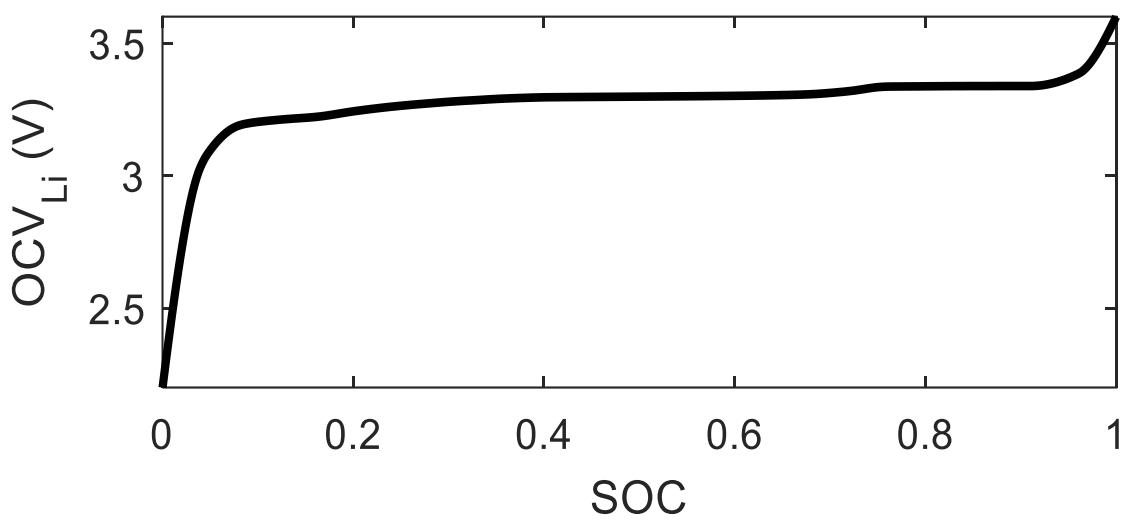

**FIGURE A.1:** OPEN-CIRCUIT VOLTAGE OF LI-ION BATTERY

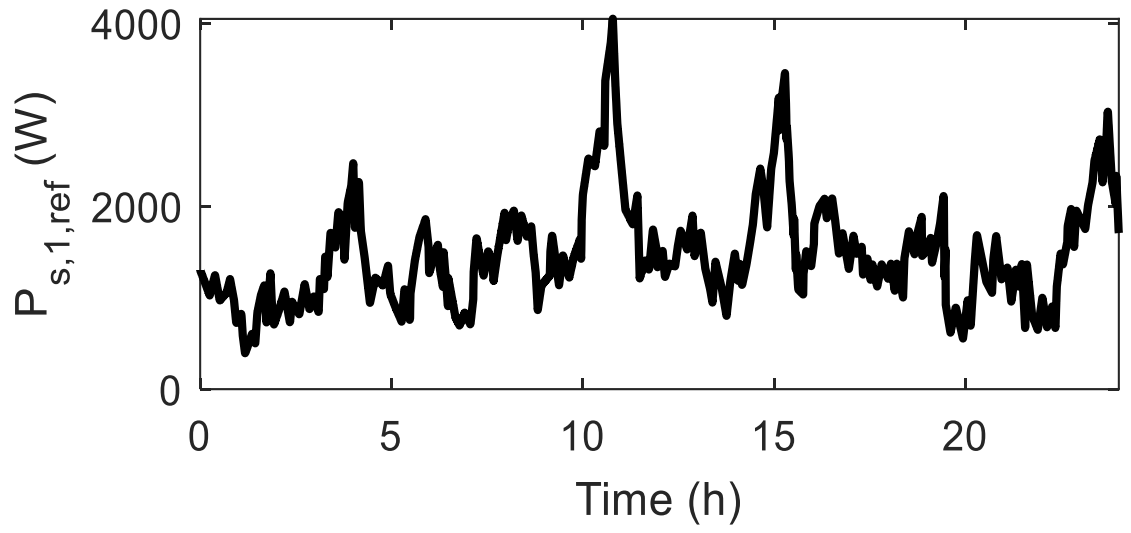

**FIGURE A.2:** SCALED SERVER RACK POWER DEMAND

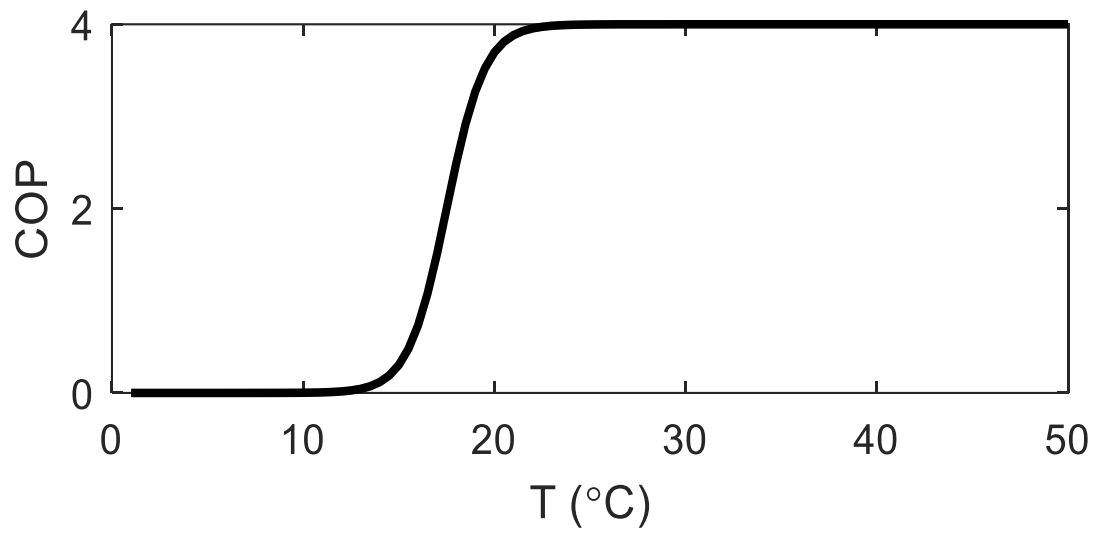

**FIGURE A.3:** COEFFICIENT OF PERFORMANCE VS. THE LOAD TEMPERATURE OF THE VCS

Defining the currents for the converters, we have:

$$I_{in,1} = N_{par,PV} I_{PV} \tag{B.1}$$

$$I_{out,1} = \frac{1}{R_{b,2}}\left(V_{out,1} - V_{bus,1}\right) \tag{B.2}$$

$$I_{in,2} = I_{Li} \tag{B.3}$$

$$I_{out,2} = \frac{1}{R_{b,2}}\left(V_{out,2} - V_{bus,1}\right) \tag{B.4}$$

$$I_{in,3} = \frac{1}{R_{b,2}}\left(V_{bus,1} - V_{in,3}\right) \tag{B.5}$$

$$I_{out,3} = \frac{1}{R_{b,2}}\left(V_{out,3} - V_{bus,2}\right) \tag{B.6}$$

For the $j = [4..6]$ converters, there is a load resistance $R_{load}$ that captures the conversion of electrical energy to power the respective load. This is shown in Equations (B.7)-(B.8).

$$I_{in,j} = \frac{1}{R_{b,2}}\left(V_{bus,2} - V_{in,j}\right) \tag{B.7}$$

$$I_{out,j} = \frac{V_{out,j}}{R_{load,j}} \tag{B.8}$$

Connecting the voltages of the converters to the battery and PV voltages, we get:

$$V_{PV} = V_{in,1} \tag{B.9}$$

$$V_{Li} = V_{in,2} \tag{B.10}$$

Table B.1 provides the mass flow rates for individual sections of the system in terms of the maximum mass flow rate, $\dot{m}_{tot}$. Table B.2 matches the general terms of Equation (15) to the specific ones from Figure 9 for each server rack. Tables B.3, B.4, and B.5 depict the actual terms used for each control volume as relating to Figure 9 and Equations (16)-(18).

## C. Controller Map

This section provides the nine discrete modes of the model free controller, shown in Table C.1.

**Table B.1:** COMPONENT-WISE MASS FLOW RATES

| Mass flow rates number | Amount of mass flow rate |
|---|---|
| $\dot{m}_1$ | $\beta\dot{m}_{tot}$ |
| $\dot{m}_2$ | $(1-\beta)\dot{m}_{tot}$ |
| $\dot{m}_3$ | $\gamma(1-\beta)\dot{m}_{tot}$ |
| $\dot{m}_4$ | $(1-\gamma)(1-\beta)\dot{m}_{tot}$ |
| $\dot{m}_5$ | $\alpha\dot{m}_{tot}$ |
| $\dot{m}_6$ | $(1-\alpha)\dot{m}_{tot}$ |

**Table B.2:** DEFINITIONS OF EACH SERVER'S DYNAMIC VARIABLES AS CORRELATED WITH EQUATION (15)

| | Server rack | |
|---|---|---|
| | $i = 1$ | $i = 2$ |
| $\mathbb{T}_{a,l,i}$ | $T_{sr,1}$ | $T_{sr,2}$ |
| $\mathbb{T}_{a,r,i}$ | $T_{sr,2}$ | $T_{sr,3}$ |
| $\mathbb{T}_{a,c,i}$ | $T_{sr,5}$ | $T_{sr,7}$ |

**Table B.3:** DEFINITIONS OF EACH SERVER ROOM AIR CV'S DYNAMIC VARIABLES AND PARAMETERS AS CORRELATED WITH EQUATION (16)

| | Server room air | | | | | | | |
|---|---|---|---|---|---|---|---|---|
| | $j = 1$ | $j = 2$ | $j = 3$ | $j = 4$ | $j = 5$ | $j = 6$ | $j = 7$ | $j = 8$ |
| $\dot{m}_{sr,1,j}$ | $\dot{m}_1$ | $\dot{m}_3$ | $\dot{m}_4$ | $\dot{m}_1$ | $\dot{m}_2$ | $\dot{m}_3$ | $\dot{m}_4$ | $\dot{m}_4$ |
| $\dot{m}_{sr,2,j}$ | 0 | 0 | 0 | $\dot{m}_2$ | 0 | $\dot{m}_4$ | 0 | 0 |
| $\dot{m}_{sr,3,j}$ | $\dot{m}_1$ | $\dot{m}_3$ | $\dot{m}_4$ | $\dot{m}_{tot}$ | $\dot{m}_2$ | $\dot{m}_2$ | $\dot{m}_4$ | $\dot{m}_4$ |
| $\mathbb{T}_{sr,1,j}$ | $T_{p,1}$ | $T_{p,2}$ | $T_{p,3}$ | $T_{sr,1}$ | $T_{sr,6}$ | $T_{sr,2}$ | $T_{sr,8}$ | $T_{sr,3}$ |
| $\mathbb{T}_{sr,2,j}$ | 0 | 0 | 0 | $T_{sr,5}$ | 0 | $T_{sr,7}$ | 0 | 0 |
| $\mathbb{T}_{sr,3,j}$ | $T_{s,1}$ | $T_{s,2}$ | $T_\infty$ | 0 | $T_{s,1}$ | 0 | $T_{s,2}$ | 0 |
| $\mathbb{T}_{sr,4,j}$ | $T_\infty$ | $T_{s,1}$ | $T_{s,2}$ | $T_\infty$ | $T_\infty$ | 0 | $T_\infty$ | $T_\infty$ |

**Table B.4:** DEFINITIONS OF EACH CCU AIR CV'S DYNAMIC VARIABLES AND PARAMETERS AS CORRELATED WITH EQUATION (17)

| | CCU | | | |
|---|---|---|---|---|
| | $k = 1$ | $k = 2$ | $k = 3$ | $k = 4$ |
| $\dot{m}_{ccu,1,k}$ | $\dot{m}_{tot}$ | $\dot{m}_5$ | $\dot{m}_6$ | $\dot{m}_{tot}$ |
| $\dot{m}_{ccu,2,k}$ | 0 | 0 | $\dot{m}_5$ | 0 |
| $\dot{m}_{ccu,3,k}$ | $\dot{m}_5$ | $\dot{m}_5$ | $\dot{m}_{tot}$ | $\dot{m}_{tot}$ |
| $\dot{m}_{ccu,4,k}$ | $\dot{m}_6$ | 0 | 0 | 0 |
| $\mathbb{T}_{ccu,1,k}$ | $T_{sr,4}$ | $T_\infty$ | $T_{ccu,1}$ | $T_{ccu,3}$ |
| $\mathbb{T}_{ccu,2,k}$ | 0 | 0 | $T_{ccu,2}$ | 0 |
| $P_{cool,k}$ | 0 | 0 | 0 | $X_{vcs}COP$ |

**Table B.5:** DEFINITIONS OF EACH PLENUM AIR CV'S DYNAMIC VARIABLES AND PARAMETERS AS CORRELATED WITH EQUATION (18)

| | Plenum | | | |
|---|---|---|---|---|
| | $n=1$ | $n=2$ | $n=3$ | $n=4$ |
| $\dot{m}_{p,1,n}$ | $\dot{m}_{tot}$ | $\dot{m}_2$ | $\dot{m}_4$ | $\dot{m}_{p,1,n}$ |
| $\dot{m}_{p,2,n}$ | $\dot{m}_1$ | $\dot{m}_3$ | $\dot{m}_4$ | $\dot{m}_{p,2,n}$ |
| $\dot{m}_{p,3,n}$ | $\dot{m}_2$ | $\dot{m}_4$ | 0 | $\dot{m}_{p,3,n}$ |
| $\mathbb{T}_{p,1,k}$ | $T_{ccu,4}$ | $T_{p,1}$ | $T_{p,2}$ | $\mathbb{T}_{p,1,k}$ |

**Table C.1:** DIFFERENT MODES OF THE MODEL-FREE CONTROLLER

| | | | |
|---|---|---|---|
| **High (Rack Temperature)** | **Mode 5**<br>$D_1$: $\alpha_1 = P_{PV,k} - P_{PV,k-1}$, $\beta_1 = V_{PV,k} - V_{PV,k-1}$<br>$D_2$: $\alpha_2 = SOC_{ref,k} - SOC_k$, $\beta_2 = D_{2,k} - D_{2,k-1}$<br>$D_3$: $D_3 = 0.5$<br>$D_4$: $D_4 = 0$<br>$D_5$: $D_5 = 0$<br>$D_6$: $D_6 = 0$ | **Mode 4**<br>$D_1$: $\alpha_1 = P_{PV,k} - P_{PV,k-1}$, $\beta_1 = V_{PV,k} - V_{PV,k-1}$<br>$D_2$: $D_2 = 0.9$<br>$D_3$: $D_3 = 0.5$<br>$D_4$: $\alpha_4 = T_{s,1,ref,k} - T_{s,1,k}$, $\beta_4 = D_{4,k} - D_{4,k-1}$<br>$D_5$: $\alpha_5 = P_{s,1,ref,k} - P_{s,1,k}$, $\beta_5 = D_{5,k} - D_{5,k-1}$<br>$D_6$: $\alpha_6 = P_{s,1,ref,k} - P_{s,1,k}$, $\beta_6 = D_{6,k} - D_{6,k-1}$ | **Mode 3**<br>$D_1$: $D_1 = 0$<br>$D_2$: $\alpha_2 = SOC_{ref,k} - SOC_k$, $\beta_2 = D_{2,k} - D_{2,k-1}$<br>$D_3$: $D_3 = 0.5$<br>$D_4$: $D_4 = 1$<br>$D_5$: $\alpha_5 = P_{s,1,ref,k} - P_{s,1,k}$, $\beta_5 = D_{5,k} - D_{5,k-1}$<br>$D_6$: $\alpha_6 = P_{s,1,ref,k} - P_{s,1,k}$, $\beta_6 = D_{6,k} - D_{6,k-1}$ |
| **Acceptable (Rack Temperature)** | **Mode 6**<br>$D_1$: $\alpha_1 = P_{PV,k} - P_{PV,k-1}$, $\beta_1 = V_{PV,k} - V_{PV,k-1}$<br>$D_2$: $\alpha_2 = SOC_{ref,k} - SOC_k$, $\beta_2 = D_{2,k} - D_{2,k-1}$<br>$D_3$: $D_3 = 0.5$<br>$D_4$: $D_4 = 0$<br>$D_5$: $D_5 = 0$<br>$D_6$: $D_6 = 0$ | **Mode 1**<br>$D_1$: $\alpha_1 = P_{PV,k} - P_{PV,k-1}$, $\beta_1 = V_{PV,k} - V_{PV,k-1}$<br>$D_2$: $D_2 = 0.9$<br>$D_3$: $D_3 = 0.5$<br>$D_4$: $D_4 = 0.5$<br>$D_5$: $\alpha_5 = P_{s,1,ref,k} - P_{s,1,k}$, $\beta_5 = D_{5,k} - D_{5,k-1}$<br>$D_6$: $\alpha_6 = P_{s,1,ref,k} - P_{s,1,k}$, $\beta_6 = D_{6,k} - D_{6,k-1}$ | **Mode 2**<br>$D_1$: $D_1 = 0$<br>$D_2$: $\alpha_2 = SOC_{ref,k} - SOC_k$, $\beta_2 = D_{2,k} - D_{2,k-1}$<br>$D_3$: $D_3 = 0.5$<br>$D_4$: $D_4 = 1$<br>$D_5$: $\alpha_5 = P_{s,1,ref,k} - P_{s,1,k}$, $\beta_5 = D_{5,k} - D_{5,k-1}$<br>$D_6$: $\alpha_6 = P_{s,1,ref,k} - P_{s,1,k}$, $\beta_6 = D_{6,k} - D_{6,k-1}$ |
| **Low (Rack Temperature)** | **Mode 7**<br>$D_1$: $\alpha_1 = P_{PV,k} - P_{PV,k-1}$, $\beta_1 = V_{PV,k} - V_{PV,k-1}$<br>$D_2$: $\alpha_2 = SOC_{ref,k} - SOC_k$, $\beta_2 = D_{2,k} - D_{2,k-1}$<br>$D_3$: $D_3 = 0.5$<br>$D_4$: $D_4 = 0$<br>$D_5$: $D_5 = 0$<br>$D_6$: $D_6 = 0$ | **Mode 8**<br>$D_1$: $\alpha_1 = P_{PV,k} - P_{PV,k-1}$, $\beta_1 = V_{PV,k} - V_{PV,k-1}$<br>$D_2$: $D_2 = 0.9$<br>$D_3$: $D_3 = 0.5$<br>$D_4$: $D_4 = 0$<br>$D_5$: $\alpha_5 = P_{s,1,ref,k} - P_{s,1,k}$, $\beta_5 = D_{5,k} - D_{5,k-1}$<br>$D_6$: $\alpha_6 = P_{s,1,ref,k} - P_{s,1,k}$, $\beta_6 = D_{6,k} - D_{6,k-1}$ | **Mode 9**<br>$D_1$: $D_1 = 0$<br>$D_2$: $\alpha_2 = SOC_{ref,k} - SOC_k$, $\beta_2 = D_{2,k} - D_{2,k-1}$<br>$D_3$: $D_3 = 0.5$<br>$D_4$: $D_4 = 1$<br>$D_5$: $\alpha_5 = P_{s,1,ref,k} - P_{s,1,k}$, $\beta_5 = D_{5,k} - D_{5,k-1}$<br>$D_6$: $\alpha_6 = P_{s,1,ref,k} - P_{s,1,k}$, $\beta_6 = D_{6,k} - D_{6,k-1}$ |
| | **Low (Li-ion battery SOC)** | **Acceptable (Li-ion battery SOC)** | **High (Li-ion battery SOC)** |